\documentclass[a4paper,fleqn,usenatbib]{mnras}

\usepackage{newtxtext,newtxmath}

\usepackage{graphicx}	
\usepackage{amsmath}	
\usepackage{amssymb}	
\usepackage[utf8]{inputenc}
\usepackage{graphics,epsfig,ulem,times}	
\usepackage{pgfplots}

\usepackage{tikz}				
\usetikzlibrary{positioning}
\usetikzlibrary{external}
\usepackage{array}
\newcolumntype{C}[1]{>{\centering\arraybackslash}p{#1}}
\newcolumntype{?}{!{\vrule width 2.5\arrayrulewidth}}
\usepackage{multirow, makecell}

\newcommand{\hMpc}{\ensuremath{~h^{-1}\text{Mpc}}}
\newcommand{\hGpc}{\ensuremath{~h^{-1}\text{Gpc}}}
\newcommand{\hMpcSquare}{\ensuremath{~(h^{-1}\text{Mpc})^{2}}}
\newcommand{\hMpcCube}{\ensuremath{~(h^{-1}\text{Mpc})^{-3}~}}
\newcommand{\hMpcInv}{\ensuremath{~h\text{Mpc}^{-1}}}
\newcommand{\Msun}{\ensuremath{~h^{-1}\text{M}_\odot}}

\newcommand{\changed}[1]{{#1}}

\title[BAO reconstruction using biased tracers]{Reconstructing the baryon acoustic oscillations using biased tracers}

\author[J. Birkin et al.]{
Jack Birkin,$^{1}$\thanks{E-mail: jack.birkin@durham.ac.uk}
Baojiu Li,$^{1}$
Marius Cautun$^{1}$
and Yanlong Shi$^{2,3}$
\\
$^{1}$Institute for Computational Cosmology, Department of Physics, Durham University, South Road, Durham, DH1 3LE, UK\\
$^{2}$Department of Astronomy, University of Science and Technology of China, Hefei 230026, Anhui, China\\
$^{3}$Department of Physics, California Institute of Technology, Pasadena, CA 91125, USA\\
}

\pubyear{2018}

\begin{document}

\setlength{\parskip}{0pt}
\label{firstpage}
\pagerange{\pageref{firstpage}--\pageref{lastpage}}
\maketitle

\begin{abstract}
The reconstruction of the initial conditions of the Universe is an important topic in cosmology, particularly in the context of sharpening the measurement of the baryon acoustic oscillation (BAO) peak. Nonlinear reconstruction algorithms developed in recent years, when applied to late-time matter fields, can recover to a substantial degree the initial density distribution, however, when applied to sparse tracers of the matter field, the performance is poorer. In this paper we \changed{apply the Shi et al. non-linear reconstruction method to biased tracers in order to establish what factors affect the reconstruction performance}. We find that grid resolution, tracer number density and mass assignment scheme all have a significant impact on the performance \changed{of our reconstruction method}, with triangular-shaped-cloud (TSC) mass assignment and a grid resolution of ${\sim}1{-}2h^{-1}$ Mpc being the optimal choice. We also \changed{show that our method can be easily adapted to include} generic tracer biases up to quadratic order in the reconstruction formalism. \changed{Applying the reconstruction to halo and galaxy samples with a range of tracer number densities, we find that the linear bias is by far the most important bias term, while including nonlocal and nonlinear biases only leads to marginal improvements on the reconstruction performance.} Overall, including bias in the reconstruction substantially improves the recovery of BAO wiggles, down to $k\sim0.25~h\text{Mpc}^{-1}$ for tracer number densities between $2\times10^{-4}$ and $2\times10^{-3}~(h^{-1}\text{Mpc})^{-3}$.

\end{abstract}




\section{Introduction}
\label{sec:intro}

The measurement of distances on large scales is one of the biggest cosmological challenges, with crucial implications for our understanding of the Universe. Distance measurement techniques usually rely on observables which can be predicted theoretically, such as the peak luminosity of a Type Ia supernova light curve, which can be used as a 'standard candle', the Tully-Fisher relation which links the angular velocity of a spiral galaxy with its intrinsic luminosity, or the relationship between the pulsation period of a Cepheid variable and its luminosity. In this work we are concerned with the baryon acoustic oscillations (BAO) which result from the propagation of sound waves in the baryon-photon    fluid prior to recombination that imprints a characteristic length scale on the large-scale structure of the Universe \citep[][]{Cole:2005sx,Eisenstein:2005su}, providing us with a so-called 'standard ruler' that can be used to measure (angular diameter) distances. In the current standard cosmological model, this primordial baryon-photon fluid is highly homogeneous, with tiny density fluctuations. Overdense regions are subject to a higher pressure, causing the fluid to propagate outwards from their centres. These `ripples' propagate at speeds of order $c$ until recombination, when the photons decouple from the baryons, leaving a shell of baryonic matter with a radius determined by the distance travelled since recombination. Since the Universe contains many of these small overdensities, these shells overlap and interfere, and the result is that matter clusters with a characteristic scale -- the BAO scale. Statistically, the BAO manifests as a peak in the matter correlation function, $\xi(r)$, at $r\sim100~h^{-1}$Mpc, or as a series of oscillations in the matter power spectrum $P(k)$, which is the Fourier transform of the correlation function \citep{2007ApJ...664..675E}. Large scale surveys of the past, present and future (e.g. \textsc{sdss} \citep{2017MNRAS.470.2617A}, 
\textsc{lsst} \citep{2008arXiv0805.2366I}, \textsc{euclid} \citep{2011arXiv1110.3193L} and \textsc{desi} \citep{2016arXiv161100036D}) map the distribution of matter on large scales, allowing us to measure \changed{the angular diameter distance $d_A$ and the Hubble parameter $H$ as a function of redshift} and \changed{thus} map the cosmic expansion history. This can be particularly effective when combined with data from type Ia supernovae and the cosmic microwave background (CMB) \citep[see e.g.][]{2015PhRvD..92l3516A}.

As the number of large-scale galaxy surveys grows, so does our ability to map the Universe to higher redshifts and measure the size of the BAO features to high precision. However, as the majority of our observations are at a relatively low redshift (as an example, \textsc{desi} will target quasars up to $z\sim3.5$ for Ly-$\alpha$ forest absorption; \citealt{2016arXiv161100036D}), the BAO peak in the correlation function is weakened and broadened due to non-linear structure formation. As a result, measurements of the BAO scale are less precise and the constraints that we can place on our cosmological model are weaker \citep{2017ApJ...841L..29W}.

A common solution to this problem is to `reverse' the evolution of the Universe and recover the distribution of matter at early times, before non-linear evolution weakened the BAO signal. This process, known as `reconstruction', dates back long before the discovery of the BAO to the work of \cite{1989ApJ...344L..53P}, who attempted to predict the trajectories of Local Group galaxies by applying the principle of least action. \cite{1992MNRAS.254..315W} proposed the 'Gaussianization' method for reconstruction, which is centred on the assumption that gravitational evolution preserves the rank order of the initial density field. \cite{1997MNRAS.285..793C} introduced the Path Interchange Zel'dovich Approximation (PIZA) method, assuming that the initial conditions are homogeneous and swapping pairs of particles in the final distribution until the total action in the evolution between the initial and final states is minimised. 
\changed{Recently, forward reconstruction models of the initial conditions have gained a lot of attention \citep[e.g.][]{Kitaura2008,Jasche2013,Wang2014,Lavaux2016}. These employ efficient Monte Carlo sampling of the initial power spectrum and phases, which is then non-linearly evolved to low redshift and compared against observations. While such methods can recover the initial conditions down to scales of a few $\rm{Mpc}$, they come at the expense of a large computational cost and complex modelling of bias and redshift space distortions \citep[e.g. see][]{Jasche2018}.}

It was first shown by \cite{2007ApJ...664..675E} that the weakening of the BAO signal is reversible, by suggesting that one can use linear theory to determine the velocity field from the density field, and subsequently reverse the gravitational flow of objects to (almost) recover their initial positions. Even with this relatively simple argument, \citeauthor{2007ApJ...664..675E} have shown that the reconstruction procedure can considerably enhance the BAO peak in the correlation function, or equivalently the oscillations in the power spectrum. \changed{\cite{2012MNRAS.427.2132P} provided the first application of reconstruction to survey data, finding a $\sim$50$\%$ reduction in the uncertainty associated to the BAO scale measurement in the SDSS Data Release 7 \citep[see also][for more recent examples]{2015PhRvD..92l3516A,2017MNRAS.470.2617A}.}

\changed{The method mentioned above makes use of the Zel'dovich approximation which is accurate down to quasi-linear scales. More recently proposed techniques, including our own, \citep[see][]{PhysRevD.97.023505}, 
extend into the non-linear regime and can therefore recover information from the initial conditions on scales of several $\rm{Mpc}$. For example, a method which is closely related to ours is the Monge-Ampe\'re-Kantorovich technique of \cite{2002Natur.417..260F}, \cite{2003MNRAS.346..501B} and \cite{2006MNRAS.365..939M}. These works presented and subsequently built on the idea that reconstruction can be treated as an example of the optimal mass transportation problem. We will see in Section~\ref{sec:method} that our method begins with the same basic principles and assumptions. More recent non-linear methods include, but are not limited to:} the nonlinear isobaric reconstruction technique of \citet[][see also \citealt{2017ApJ...841L..29W}]{2017PhRvD..96l3502Z}, 
the iterative technique described in \citet{2017PhRvD..96b3505S} and the multigrid relaxation method proposed by \cite{PhysRevD.97.023505}, the latter of which this work will build on.
All of the aforementioned methods have been shown to be capable of recovering the initial conditions on intermediate to non-linear scales when applied to a late-time matter field, and, in the case of the \citeauthor{2017PhRvD..96l3502Z} method, a late-time halo field \citep{2017ApJ...847..110Y}. For example, \cite{2017ApJ...841L..29W} showed that isobaric reconstruction could significantly recover the BAO signal from the matter field.

When a reconstruction method is applied to a tracer field, such as a halo or a galaxy field, an additional complication is the biasing between the tracer and underlying matter fields. Dark matter halos and galaxies, for example, are known to be biased tracers, i.e. their density fields are different from the matter density field. Reconstruction directly from the former, therefore, can lead to errors in the recovery of the initial matter distribution and hence the position and width of the BAO peaks. This issue has been discussed in, e.g., \citet{Wang:2018ika} which shows the non-negligible effect of halo bias on the reconstruction of BAO wiggles.

In this paper, we extend the reconstruction method of \citet{PhysRevD.97.023505} to accommodate biased tracers and develop it further to include up to quadratic order bias schemes. We then investigate how including these bias terms impacts on the reconstruction performance and results. Furthermore, we also study the effect of grid size, mass assignment scheme and tracer number density on the reconstruction performance. We do so for both halo and galaxy distributions with varying number densities.

The paper is organised as following: In Section~\ref{sec:method} we give a brief review of the \citet{PhysRevD.97.023505} reconstruction method and describe the extension for including biased tracers. In Section~\ref{sec:simulations} we detail the simulations used, along with the halo and galaxy fields used in this work. Section~\ref{sec:results} contains the main results, including tests of the impacts of a number of factors that can physically or numerically affect reconstruction performance, and the effects of including tracer biases up to the quadratic order. We then show how our biased reconstruction method can help improve the measurements of BAO wiggles from the tracer power spectra. Finally, Section~\ref{sec:conclusions} presents a summary of the findings of this paper, conclusions and discussions of possible future work.

\section{Reconstruction Method}
\label{sec:method}

\subsection{The reconstruction equation}

We assume that the initial Lagrangian position $\textbf{q}$ of a particle can be mapped to its final Eulerian position $\textbf{x}$ by the gradient of a 'displacement potential' $\Theta$, i.e.
\begin{equation}
	\mathbf{q} = \nabla_{\mathbf{x}}\Theta(\mathbf{x}).
	\label{eq:11}
\end{equation}
This is valid on large scales where stream crossing has not occurred\footnote{Note that, due to the hierarchical nature of structure formation in $\Lambda$CDM cosmology, stream crossing, i.e., particles crossing the trajectories of each other, is inevitable on small enough scales. Therefore, the assumption of no stream crossing is good only on large enough scales. We shall see later that this means that the reconstruction method is less accurate on smaller scales.}.   
Note that Eq.~(\ref{eq:11}) also assumes
that there is no curl component in the relation between $\mathbf{q}$ and $\mathbf{x}$. The absence of stream crossing also implies mass conservation in a given volume element:
\begin{equation}\label{eq:mass_cons}
\rho({\bf x}){\rm d}^3{\bf x} = \rho({\bf q}){\rm d}^3{\bf q} \approx \bar{\rho}{\rm d}^3{\bf q},
\end{equation}
where ${\rm d}^3{\bf q}$ and ${\rm d}^3{\bf x}$ are the volume elements at the initial and final times, respectively, and $\rho({\bf q})$ and $\rho({\bf x})$ are the densities of the corresponding volume elements. The Universe is almost homogeneous at early times, however, and we can therefore assume that $\rho({\bf q})\approx\bar{\rho}$, where $\bar{\rho}$ is the mean matter density.

Using Eq.~(\ref{eq:11}), Eq.~\eqref{eq:mass_cons} 
can be rearranged to obtain
\begin{equation}\label{eq:12}
\text{det}\left[\nabla^i\nabla_j\Theta(\bf{x})\right] = \dfrac{\rho\left(\bf{x}\right)}{\overline{\rho}} \equiv 1+\delta\left({\bf x}\right),
\end{equation}
where $i$, $j$ = 1,2,3 represent the 3 cartesian coordinates and $\delta$ is the density contrast. The LHS of Eq.~(\ref{eq:12}) represents a Jacobian matrix comprising the derivatives of the three components of $\textbf{q}$ with respect to the three components of $\textbf{x}$. \cite{2002Natur.417..260F} found the solution to Eq.~(\ref{eq:12}) by treating reconstruction as an 'optimisation problem' and finding the arrangement of particles which minimises a 'cost function'. Similar to the PIZA method, an algorithm is used to swap particles in the final distribution until the optimal arrangement is obtained.

In this work we follow the new and efficient method developed in \cite{PhysRevD.97.023505}, and recast Eq.~(\ref{eq:12}) into a nonlinear elliptical partial differential equation (PDE) which can be solved numerically. The result is
\begin{equation}
	\dfrac{1}{6}\left(\nabla^2\Theta\right)^3 - \dfrac{1}{2}\nabla^i\nabla_j\Theta\nabla^j\nabla_i\Theta\nabla^2\Theta + \dfrac{1}{3}\nabla^i\nabla_j\Theta\nabla^j\nabla_k\Theta\nabla^k			\nabla_i\Theta = \dfrac{\rho(\bf{x})}{\overline{\rho}},
	\label{eq:13}
\end{equation} 
where we have used the Einstein summation convention. We shall apply the multigrid relaxation technique to solve Eq.~(\ref{eq:13}) for $\Theta$, but for numerical implementation it is essential to split $\nabla^i\nabla_j\Theta$ into a diagonal part and a traceless part (see \citealt{PhysRevD.97.023505} for a more detailed description of the numerical algorithm) as follows
\begin{equation}
	\nabla^i\nabla_j\Theta \equiv \dfrac{1}{3}\delta_{ij}\nabla^2\Theta + \bar{\nabla}^i\bar{\nabla}_j\Theta,
	\label{eq:14}
\end{equation}
which can be regarded as a definition of the barred derivative $\bar{\nabla}^i$. Inserting Eq.~\eqref{eq:14} into Eq.~\eqref{eq:13} gives
\begin{equation}
	(\nabla^2\Theta)^3 - \dfrac{9}{2}\bar{\nabla}^i\bar{\nabla}_j\Theta\bar{\nabla}^j\bar{\nabla}_i\Theta\nabla^2\Theta + 9\bar{\nabla}^i\bar{\nabla}_j\Theta\bar{\nabla}^j				\bar{\nabla}_k\Theta\bar{\nabla}^k\bar{\nabla}_i\Theta - 27(1+\delta) = 0,
	\label{eq:15}
\end{equation}
which we will refer to as the {\it reconstruction equation} from now on.

In \cite{PhysRevD.97.023505}, this method was studied in the context of reconstruction from a late-time matter density field, where it was shown to be capable of recovering \changed{the sharpness of the first five BAO peaks}. However, cosmological observations do not usually provide us with the 3D matter density fields, but instead catalogues of tracers of the large-scale structure, such as galaxies, clusters, quasars or 21cm intensities. These tracers are biased, i.e., $\delta_{\rm tracer}({\bf x})\equiv n_{\rm tracer}({\bf x})/\bar{n}_{\rm tracer}-1$, where $n_{\rm tracer}({\bf x})$ is the number density of the tracer type at ${\bf x}$ and $\bar{n}_{\rm tracer}$ is its mean value, is generally not equal to the matter density contrast $\delta({\bf x})\equiv\rho({\bf x})/\bar{\rho}-1$. In the simplest case, a constant linear bias $b_1$ applies, where $\delta_{\rm tracer}=b_1\delta$, but this usually works only on very large scales, while in general the bias effects can be more complicated and include nonlinear and nonlocal terms \citep[][see e.g. \citealt{Desjacques:2016bnm} for a comprehensive review]{Fry:1992vr,2012PhRvD..85h3509C}. 
Clearly, as the reconstruction algorithm described above requires $\delta({\bf x})$, while observations give $\delta_{\rm tracer}({\bf x})$, the bias needs to be included in the reconstruction procedure.

As we shall now see, our method can be naturally extended to include the effects of nonlinear and nonlocal biases. For simplicity, here we consider these bias parameters up to second order, in which case the matter and tracer density contrasts are related by
\begin{equation}
	\delta_h = b_1\delta + \dfrac{b_2}{2}\delta^2 + \gamma_2\mathcal{G}_2,
	\label{eq:29}
\end{equation}
where $b_1$ is the linear bias, $b_2$ is the quadratic bias, $\gamma_2$ is a nonlocal bias parameter, and $\delta_h$ replaces $\delta_{\rm tracer}$ to make the notation more compact, representing the number density contrast of halos, although this could be interchanged with any tracer type. The nonlocal bias term in Eq.~\eqref{eq:29} can be expressed as \citep{2012PhRvD..85h3509C}:
\vspace{-3ex}
\begin{equation}
	\mathcal{G}_2 = \nabla^i\nabla_j\Phi_v\nabla^j\nabla_i\Phi_v - \left(\nabla^2\Phi_v\right)^2,
	\label{eq:30}
\end{equation}
where $\Phi_v$ is the the velocity potential, which in the Zel'dovich approximation is related to the displacement field by
\begin{equation}
	\Psi(\textbf{q}) = \textbf{x}(\textbf{q}) - \textbf{q} = -\nabla\Phi_v.
	\label{eq:31}
\end{equation}
As $\Psi$ can be expressed as a derivative of $\Theta$, $\nabla^2\Phi_v$ and $\nabla^i\nabla_j\Phi_v$ can be written in terms of second-order derivatives of $\Theta$. This suggests a way to include nonlocal bias in a slightly modified version of the reconstruction equation, Eq.~(\ref{eq:15}). To see this, let us note
\begin{equation}
	\nabla^i\nabla_j\Phi_v = \nabla^i(q_j-x_j) = \nabla^i\nabla_j\Theta - \delta^i_{\ j},
	\label{eq:32}
\end{equation}
and
\begin{equation}
	\nabla^2\Phi_v = \nabla^2\Theta - 3.
	\label{eq:33}
\end{equation}
Substituting Eq.~(\ref{eq:32}) and Eq.~(\ref{eq:33}) into Eq.~(\ref{eq:29}) gives
\begin{equation}\label{eq:34}
\delta = \dfrac{\delta_h}{b_1} - \dfrac{b_2}{2b_1^3}\delta_h^2 - \dfrac{\gamma_2}{b_1}\left(\nabla^i\nabla_j\Theta\nabla^j\nabla_i\Theta + 4\nabla^2\Theta - \left(\nabla^2\Theta\right)^2 - 6\right),
\end{equation}
in which the second term on the right-hand side is obtained by approximating $\delta\approx b_1^{-1}\delta_h$. One can then replace the $\delta$ in Eq.~\eqref{eq:15} using Eq.~\eqref{eq:34} to derive a modified reconstruction equation, which is still a PDE for $\Theta$ but which is now sourced by $\delta_h$ (the directly observable quantity) rather than $\delta$. 
The resulting \textit{modified reconstruction equation}, which is a more general version of Eq.~(\ref{eq:15}), is given by
\begin{equation}
\begin{split}\label{eq:rec_eqn_bias}
(\nabla^2\Theta)^3 - \dfrac{9}{2}\bar{\nabla}^i\bar{\nabla}_j\Theta\bar{\nabla}^j\bar{\nabla}_i\Theta\nabla^2\Theta + 9\bar{\nabla}^i\bar{\nabla}_j\Theta\bar{\nabla}^j\bar{\nabla}_k\Theta\bar{\nabla}^k\bar{\nabla}_i\Theta - \\
27\left[1+\dfrac{\delta_h}{b_1} - \dfrac{b_2}{2b_1^3}\delta_h^2 - \dfrac{\gamma_2}{b_1}\left(\nabla^i\nabla_j\Theta\nabla^j\nabla_i\Theta + 4\nabla^2\Theta - \left(\nabla^2\Theta\right)^2 - 6\right)\right] &= 0
\end{split} .
\end{equation}
This can be applied to any distribution of tracers, and reduces to the standard reconstruction equation in the case of the matter density field (i.e. by setting $b_1=1$, $\gamma_2=b_2=0$ and $\delta_h=\delta$).

\subsection{The numerical algorithm}

We solve for $\Theta$ numerically on a discrete grid, i.e., $\Theta \equiv \Theta_{i,j,k}$ where $i$, $j$ and $k$ are the indices of cells in the $x$, $y$ and $z$ directions respectively. A crucial benefit of the operator splitting in Eq.~\eqref{eq:14} is that $\nabla^2\Theta$ depends on $\Theta_{i,j,k}$, whereas $\bar{\nabla}^i\bar{\nabla}_j\Theta$ does not. This allows us to treat Eq.~(\ref{eq:15}) or \eqref{eq:rec_eqn_bias} as a cubic equation for $\nabla^2\Theta$, which can be solved for given $\bar{\nabla}_i\bar{\nabla}_j\Theta$ and $\delta$ (or $\delta_h$). From $\nabla^2\Theta$ we can then calculate $\Theta_{i,j,k}$.

It is useful to adjust the form of the reconstruction equations before solving them. Let us take Eq.~\eqref{eq:rec_eqn_bias} as an example here and below. Consider the case of an entirely uniform density field, i.e. $\delta(\textbf{x}) =$ 0. From Eq.~(\ref{eq:12}), $\text{det}[\nabla^i\nabla_j\Theta(\bf{x})] =$ 1, so the uniform solution is
\begin{equation}
	\Theta = \Theta_0 = \dfrac{1}{2}(x^2 + y^2 + z^2),
	\label{eq:16}
\end{equation}
and we can define a new variable $\theta$ as the perturbation of $\Theta$ around the uniform solution $\Theta_0$, i.e.
\begin{equation}
	\Theta \equiv \Theta_0 + \theta.
	\label{eq:17}
\end{equation}
It is advantageous to rewrite Eq.~(\ref{eq:rec_eqn_bias}) in terms of $\theta$ since our method for solving the PDE is iterative and requires an initial guess for the solution. If we are solving for $\theta$, which represents a perturbation from the uniform solution, then a natural choice for an initial guess is zero. Eq.~(\ref{eq:rec_eqn_bias}) can be recast into the following cubic equation for $\theta$
\begin{equation}
	a(\nabla^2\theta + 3)^3 + b(\nabla^2\theta + 3)^2 + c(\nabla^2\theta + 3) + d = 0,
	\label{eq:35}
\end{equation}
with coefficients
\begin{equation}
\begin{split}
	a &= 1 \\
	b &= -\dfrac{18\gamma_2}{b_1} \\
	c &= \dfrac{108\gamma_2}{b_1} - \dfrac{9}{2}\bar{\nabla}^i\bar{\nabla}_j\theta\bar{\nabla}^j\bar{\nabla}_i\theta \\
	d &= 9\bar{\nabla}^i\bar{\nabla}_j\theta\bar{\nabla}^j\bar{\nabla}_k\theta\bar{\nabla}^k\bar{\nabla}_i\theta - \\ & ~~~~~\dfrac{27}{b_1}\left(b_1 + \delta_h  - \dfrac{b_2}{2b_1^2} 		 \delta_h^2 - \gamma_2\bar{\nabla}^i\bar{\nabla}_j\theta\bar{\nabla}^j\bar{\nabla}_i\theta + 6\gamma_2\right),
	\label{eq:36}
\end{split}
\end{equation}
In practice, the quantities $\nabla^2\theta$, $c$ and $d$ in Eq.~(\ref{eq:35}) are calculated on a discretised grid ($a$ and $b$ are constants), and one should add the subscripts $i$, $j$ and $k$ to label the coordinate of the cell, but these are omitted here for brevity. 

As a cubic equation, Eq.~\eqref{eq:35} has multiple analytical solutions, meaning we need a method for establishing which solution is physical. To this end, we define the discriminant as 
\begin{equation}
	\Delta \equiv \dfrac{q^2}{4} + \dfrac{p^3}{27},
	\label{eq:20}
\end{equation}
where
\begin{subequations}
\begin{align}
	p = \dfrac{3ac - b^2}{3a^2},
	\label{eq:38a}
\end{align}
\begin{align}
	q = \dfrac{2b^3 - 9abc + 27a^2d}{27a^3},
	\label{eq:38b}
\end{align}
\end{subequations}
For $\Delta \geq 0$ the equation has a single real root, which is the physical solution, while for $\Delta < 0$ there are 3 real roots, and the physical one must change continuously as $\Delta$ crosses zero. The physical solution in each case is therefore found to be
\begin{subequations}
\begin{align}
	\nabla^2\theta &= -3 + \left[-\dfrac{q}{2}+\Delta^{\frac{1}{2}}\right]^{\frac{1}{3}} + \left[-\dfrac{q}{2}-\Delta^{\frac{1}{2}}\right]^{\frac{1}{3}} & \text{if}\ \Delta\geq0,
	\label{eq:21a}
\end{align}
\vspace{-2.5ex}
\begin{align}
	\nabla^2\theta &= -3 - \left(-\dfrac{p}{3}\right)^{\frac{1}{2}}\text{cos}\left[\dfrac{1}{3}(\sigma+2\pi)\right] & \text{if}\ \Delta<0,
	\label{eq:21b}
\end{align}
\end{subequations}
where
\begin{equation}
	\text{cos}(\sigma) \equiv \dfrac{3q}{2p}\left(\dfrac{-3}{p}\right)^{\frac{1}{2}},
	\label{eq:22}
\end{equation}
and $\sigma$ takes a value between 0 and $\pi$. 

Eq.~(\ref{eq:21a}) and Eq.~(\ref{eq:21b}) are then solved to find $\theta$ using a multigrid Gauss-Seidel technique. As previously mentioned, these two equations are discretised on mesh cells ($\theta \rightarrow \theta_{i,j,k}$). As $\theta_{i,j,k}$ is not a continuous function, the spatial derivatives such as $\nabla \theta$ have to be calculated as finite differences, e.g.,
\begin{equation}
    \nabla_x \theta = \dfrac{1}{2\ell}\left(\theta_{i+1,j,k} - \theta_{i-1,j,k}\right).
	\label{eq:23}
\end{equation}
which represents the $x$-component of the gradient of $\theta$, and where $\ell$ is the side size of a cell which is taken as cubic for simplicity. The finite difference expression in Eq.~\eqref{eq:23} is known to have a second-order accuracy, meaning that the error due to the discretisation decreases quadratically as we reduce the cell length $\ell$. We can similarly write finite-difference expressions for higher-order derivatives of $\theta$ and their products, but for brevity these are not listed here, and interested readers can find them
in \cite{PhysRevD.97.023505}. 

Upon discretisation, Eq.~(\ref{eq:21a}) and Eq.~(\ref{eq:21b}) can be written as an operator $\mathcal{L}^\ell\left[\theta_{i,j,k}\right]$:
\begin{multline}
\mathcal{L}^\ell\left[\theta_{i,j,k}\right] = \dfrac{1}{\ell^2}\left(\theta_{i+1,j,k} + \theta_{i-1,j,k} + \theta_{i,j+1,k} + \theta_{i,j-1,k}\right. + \\ \left.\theta_{i,j,k+1} +  \theta_{i,j,k-1} - 					6\theta_{i,j,k}\right) - \Sigma_{i,j,k} = 0,
	\label{eq:25}
\end{multline} 
where $\Sigma_{i,j,k}$ is a discretisation of the RHS of Eq.~(\ref{eq:21a}) or Eq.~(\ref{eq:21b}), depending on the value of $\Delta$. As mentioned above, the use of the operator splitting ensures that $\Sigma_{i,j,k}$ does not contain $\theta_{i,j,k}$, so that $\mathcal{L}^\ell\left[\theta_{i,j,k}\right]$ is effectively a linear operator of $\theta_{i,j,k}$.

The Gauss-Seidel relaxation technique can be used to iteratively update the values of $\theta_{i,j,k}$:
\begin{equation}
	\theta^{n+1}_{i,j,k} = \theta^n_{i,j,k} - \dfrac{\mathcal{L}^\ell \left[\theta^n_{i,j,k}\right]}{\partial\mathcal{L}^\ell \left[\theta^n_{i,j,k}\right] / \partial\theta^n_{i,j,k}},
	\label{eq:26}
\end{equation}
where the superscript $n$ represents the value at the $n$th iteration (remember that the use of $\theta$ instead of $\Theta$ gives us the natural choice of $\theta^0_{i,j,k} = 0$ as the initial guess for the first iteration). While Eq.~\eqref{eq:26} is a general expression for nonlinear operators $\mathcal{L}$, because $\mathcal{L}^\ell\left[\theta_{i,j,k}\right]$ is a linear operator, one can directly write $\theta^{n+1}_{i,j,k}$ analytically as
\begin{multline}
\theta_{i,j,k}^{n+1} = \frac{1}{6}\left(\theta^{n}_{i+1,j,k}+\theta^{n+1}_{i-1,j,k}+\theta^n_{i,j+1,k}+\theta^{n+1}_{i,j-1,k}\right. \\ \left.+\theta^n_{i,j,k+1}+\theta^{n+1}_{i,j1,k-1}\right)-\frac{1}{6}\ell^2\Sigma_{i,j,k},
\end{multline}
where we note that the right-hand side uses a mixture of the $n$th and $(n+1)$th iteration values of $\theta$ in neighbouring cells of cell $(i,j,k)$ -- this is because in the Gauss-Seidel method the relaxation iterations always make use of the most updated values of neighbouring cells.

We define the residual $\epsilon$ as 
\begin{equation}
	\epsilon \equiv \left[\dfrac{1}{N^3} \sum_{i,j,k=1}^{N} \left(\mathcal{L}^\ell[\theta_{i,j,k}]\right)^2\right]^{1/2},
	\label{eq:27}
\end{equation}
where $N$ is the number of cells along each axis. Provided the algorithm is stable, $\epsilon$ decrease as the number of iterations increases. Convergence is deemed to have occurred for $\epsilon < 10^{-8}$, at which point the iterations stop and $\theta$ is outputted along with $\nabla_{\mathbf{x}} \theta$. To improve the convergence, we have used the multigrid technique \citep{Press2007}, which employs a hierarchy of coarser meshes to speed up the decrease of $\epsilon$ \citep[see][for more details]{PhysRevD.97.023505}.

The method for calculating $\theta$ and $\nabla_{\mathbf{x}} \theta$ is incorporated into the \textsc{ecosmog} code \citep[see][]{2012JCAP...01..051L}, which is based on the publicly available N-body simulation code \textsc{ramses} \citep{Teyssier:2001cp}. 
This gives us the values of $\theta(\textbf{x})$ and $\nabla_{\mathbf{x}}\theta$ on a uniform ${\bf x}$-grid, from which we can calculate the corresponding ${\bf q}({\bf x})$ coordinates. Then, the displacement field, ${\bf \Psi}({\bf q}) = \textbf{x} - \textbf{q}$, represents a vector defined at an irregular set of points with coordinates ${\bf q}$ and can be used to calculate the 
reconstructed initial density field, $\delta_r$, as
\begin{equation}
	\delta_r = \nabla_{\mathbf{q}} \cdot\Psi\left({\bf q}\right),
	\label{eq:28}
\end{equation}
which we implement using the \textsc{dtfe} code \citep[][see Sec. \ref{sec:halos} for more details]{2011ascl.soft05003C, 2000A&A...363L..29S}. Note that this calculation is very similar to the use of {\sc dtfe} to compute the velocity divergence field, for which we have the velocities ${\bf v}({\bf x})$ (analogous to ${\bf \Psi}({\bf q})$) of a set of particles with known ${\bf x}$-coordinates (analogous to the ${\bf q}$-coordinates).

\section{Simulations}
\label{sec:simulations}

\subsection{Simulation details}

We adopt a $\Lambda$CDM cosmology in our simulations. The specifications of the simulations, along with their cosmological parameters, can be found in Table~\ref{tab:simdetails}. Initial conditions were generated using second-order Lagrangian perturbation theory (2LPT, the \textsc{2lpt}ic code) \citep[see][]{1998MNRAS.299.1097S} at $z_i$ = 49, which has been found to be a suitable choice of initial redshift for 2LPT initial conditions \citep{2006MNRAS.373..369C}. 
We evolve the initial conditions using the {\sc ramses} code, which uses adaptive mesh refinement (AMR) when solving the Poisson equation, meaning that the simulations begin with a uniform domain grid until the number of particles within a cell exceeds some refinement criterion (see Table~\ref{tab:simdetails}), at which point the cell is refined to achieve a higher resolution. In our case this means that a cell will refine itself when it contains 4 particles, and the resulting cells will refine themselves again when they contain 4 particles. This pattern will continue using the refinement criterion given in Table~\ref{tab:simdetails}.

\begin{table}
	\centering
	\caption{ Cosmological parameters and simulation details. The values of the density parameters, $\Omega$, correspond to those at present day.}
	\setlength\tabcolsep{3pt}
	\renewcommand{\arraystretch}{1.0}
	\begin{tabular}{@{}C{1.7cm}C{1.7cm}?C{2cm}C{2.4cm}@{}} \Xhline{3.5\arrayrulewidth}
	\multicolumn{2}{c}{Cosmological Parameters} & \multicolumn{2}{c}{Simulation Details}\\ \Xhline{3.5\arrayrulewidth}
	Parameter & Value & Parameter & Value\\ \Xhline{2.5\arrayrulewidth}
	$\Omega_m$ & 0.3072 & Particle Number & 1024$^3$\\
	$\Omega_b$ & 0.0481 & Box Size & 1\hGpc\\
	$\Omega_{\Lambda}$ & 0.6928 & Particle Mass & 7.94 $\times$ 10$^{10}$\Msun\\
	$h$ & 0.68 & Refinement & 4, 4, 4, 5, 6, 7, 8, 8... \\
	$\sigma_8$ & 0.8205 & & \\ \Xhline{3.5\arrayrulewidth}
	\end{tabular}
	\label{tab:simdetails}
\end{table}

In order to highlight the BAO signal, in what follows we shall compare the matter power spectra from a full simulation with those from a paired no-wiggle simulation, $P_{\text{nw}}$. To generate initial conditions for these two sets of simulations, we calculated the initial matter power spectra with and without the BAO signal using the transfer functions of \cite{1998ApJ...496..605E}, and used these as the input to {\sc 2lpt}ic. More information can be found in that work, although we will state here that these functions are appropriate for a high-baryon model, which is not what we are using here. However, the objective of this work is not to accurately model the BAO wiggles, but to test to what extent the reconstruction method can recover them.
The initial conditions for the paired simulations with and without BAO wiggles were generated using the same random number seeds to ensure that the corresponding initial density fields have the same random phases and only differ by the BAO features.

\subsection{Tracers of the dark matter field}

In this work we will test the reconstruction technique when starting from late-time halo density fields, in a similar way to the study of \cite{2017ApJ...847..110Y}, and late-time galaxy density fields, proceeding to examine the effects of including halo/galaxy bias in the reconstruction. Dark matter halos are a tracer of the total matter distribution, and can be used as a rough proxy for galaxies in a large scale survey. As tracers such as galaxies and dark matter halos generally have much lower number densities than the dark matter particles in an $N$-body simulation, naturally the reconstruction performance will be worse than in \cite{PhysRevD.97.023505}. However, understanding the effects of using tracers is important since the application of reconstruction to large-scale survey data involves determining the matter density field from tracers.

\subsubsection{Dark matter halos}
\label{sec:halos}

The dark matter halo catalogues used in this paper are generated using the \textsc{rockstar} halo finder \citep{2013ApJ...762..109B}. \textsc{rockstar} uses a variant of the 3D friends-of-friends method with a modified algorithm that requires a reduced number of calculations and therefore a shorter computation time. We approximate halos as spherical objects and define their boundary to be at the radius within which their mean mass density is 200 times the critical density $\rho_{\rm crit}$ of the Universe. The halo mass, i.e., the mass contained within this radius, is denoted by $M_{200c}$. Subhalos are excluded from our analysis. We apply a mass cutoff, i.e., we ignore halos with a lower mass than this cutoff, which allows us to tune our halo catalogues to a particular number density. This will be important when comparing halo and galaxy reconstructions (we will use equal number densities for these two tracer types), and when testing reconstruction using different tracer number densities. 

We calculate the linear halo bias according to
\begin{equation}
	b_1(r) = \dfrac{\xi_{hh}(r)}{\xi_{hm}(r)},
	\label{eq:39}
\end{equation}
where $\xi_{hh}(r)$ is the halo auto-correlation function and $\xi_{hm}(r)$ is the cross correlation between the halo and the dark matter distributions. Since we have chosen a standard $\Lambda$CDM cosmology, the linear bias $b_1$ is constant for large scales.
$\xi_{hh}(r)$ and $\xi_{hm}(r)$ are computed using the Correlation Utilities and Two-Point Estimates (\textsc{cute}) code \citep{2012arXiv1210.1833A}. We calculate the large-scale value of $b_1$ by using Eq.~\eqref{eq:39} to measure $b_1(r)$ at different scales, $r$, and then taking the average value in the scale range $10-70~h^{-1}$Mpc. While this gives a reasonable estimate, in Sec. \ref{sec:results} we test several other $b_1$ values around the measured value of Eq.~\eqref{eq:39}.

According to linear perturbation theory, the nonlocal bias parameter $\gamma_2$ can be calculated by \citep{2012PhRvD..85h3509C}
\begin{equation}
	\gamma_2=\dfrac{-2(b_1-1)}{7},
	\label{eq:gamma2}
\end{equation}
although we will see that this expression does not work well for both halo and galaxy reconstruction, so we also test different values of $\gamma_2$ to see which value gives the best reconstruction performance for a given tracer number density. We do the same for the nonlinear bias to quadratic order, $b_2$.

\begin{figure*}
    \centering
	\includegraphics[width=\linewidth,angle=0]{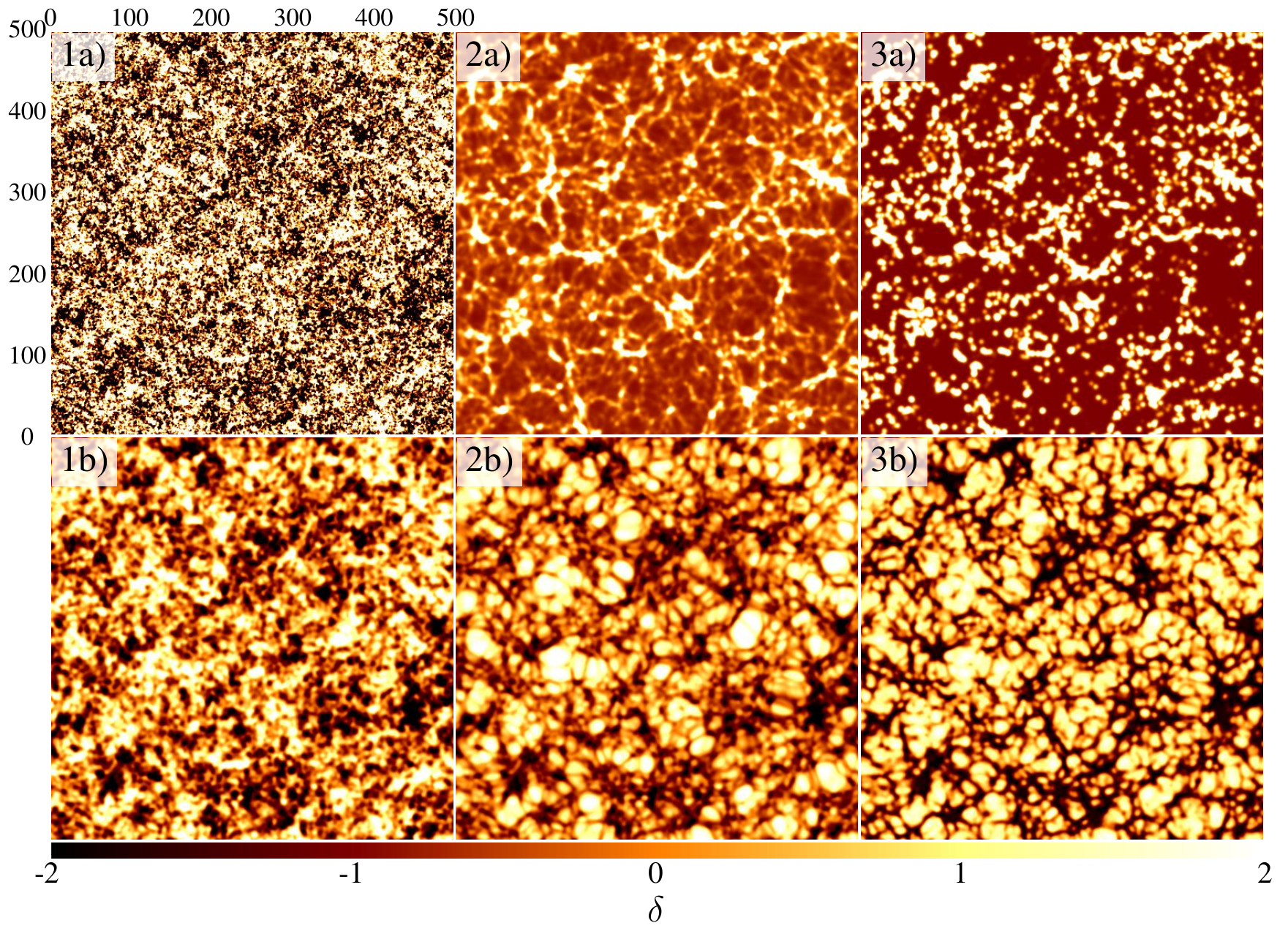}
    \vspace{-3ex}
    \caption{ Illustration of the reconstruction results. Each panel shows the same $500\times500\hMpcSquare{}$ region with $1.0\hMpc$ thickness of the simulation box. The panels 1a) and 1b) show the initial matter density contrast, $\delta({\bf x})$, linearly extrapolated to $z=0$, with 1a) corresponding to the unsmoothed density field while 1b) corresponds to the field smoothed with a spherical Gaussian filter of size, $R=2\hMpc$. Panel 2a) shows the nonlinearly evolved matter density at $z=0$, and 2b) shows the reconstructed linear density, $\delta_r({\bf x})$, from the $z=0$ dark matter distribution. Panel 3a) shows the dark matter halo number density, $\delta_h({\bf x})$, at $z=0$, and 3b) shows the reconstructed linear density from the same halo distribution. For 3a) and 3b) the halo number density is $2\times10^{-3}\hMpcCube$. 
    The density fields in 1b), 2a), 2b), 3a) and 3b) are all smoothed with the same $R = 2 \hMpc$ Gaussian filter. All six panels use the same colour scheme (see the bottom of the figure) which corresponds to the $\delta$ values shown on a linear scale between $[-2,2]$.
    }
    \label{fig:density_fields}
\end{figure*}

We compare different methods of calculating the number density field of dark matter halos, $n_{\rm halo}\left({\bf x}\right)$. The first approach consists of the Delaunay Tessellation Field Estimator (DTFE; \citealt{2000A&A...363L..29S}) method implemented in the \textsc{dtfe} code \citep{2011ascl.soft05003C}, which offers the ability to compute a continuous density field from the positions of discrete tracers. DTFE constructs a Delaunay triangulation that tessellates the entire volume with tetrahedra whose vertices are given by the distribution of tracers, which can be dark matter particles, halos or galaxies. The mass of each tracer particle is distributed among the tetrahedra which have that particle as a vertex. Then, to obtain the density on a regular grid, the mass in each tetrahedron is distributed among the grid cells which intersect that tetrahedron. The tessellation is space filling and thus all grid cells contain a non-zero mass and thus a non-zero density. The DTFE method is adaptive to the local tracer distribution: tracer particles in high number density regions typically distribute their mass to a small region around them, while tracers in low number density regions typically distribute their mass up to large distances.

In the second approach, we use the cloud-in-cell (CIC) and triangular-shaped-cloud (TSC) mass assignment schemes to calculate $n_{\rm halo}$ on the uniform grid used for reconstruction. In three dimensions, the TSC and CIC assignment schemes respectively distribute the mass of a given particle to the 27 and 8 neighbouring cells which overlap with its `cloud' \citep{Hockney_Eastwood_book}. For tracers with a low number density, and using a relatively high resolution computational grid for reconstruction, a lot of grid cells will be left with zero density. As we shall see later, this has a non-negligible impact on the reconstruction result, because TSC and CIC differ significantly from {\sc dtfe} in low-density regions, with the latter spreading masses into larger spatial regions.

In the results shown below we do not weight halos according to their mass; we treat all halos used for reconstruction as particles of equal mass. We will briefly comment on the tests and results using halo-mass-weighted reconstruction, and possible future development, in the conclusion section.

\subsubsection{Galaxies}

We build galaxy catalogues by populating halos using the Halo Occupation Distribution (HOD) method \citep{Berlind:2001xk,Zheng:2004id}, which assumes that the probability of a halo hosting one or more galaxies is dependent on the mass of the halo. Specifically,
\begin{subequations}
\begin{align}
\left\langle N_{\text{cen}}(M)\right\rangle = \dfrac{1}{2} \left[1 + \text{erf}\left(\dfrac{\text{log}M - \text{log}M_{\text{min}}}{\sigma_{\text{log}M}}\right)\right],
\end{align}
\begin{align}
\left\langle N_{\text{sat}}(M)\right\rangle = \left\langle N_{\text{cen}}(M)\right\rangle \left(\dfrac{M - M_0}{M_1}\right)^{\alpha},
\end{align}
\end{subequations}
as was suggested by \cite{2007ApJ...667..760Z}. $\left\langle N_{\text{cen}}(M)\right\rangle$ and $\left\langle N_{\text{sat}}(M)\right\rangle$ are the mean numbers of central and satellite galaxies, respectively, and erf is the error function. The number of galaxies within a halo is then a sum of the number of central and satellite galaxies. The model contains five free parameters, with our choices for these parameter values being shown in Table~\ref{tab:hodparams}.

\begin{table}
	\centering
	\caption{ The parameters of the Halo Occupation Distribution (HOD) model used to obtain galaxy catalogues. We use three different $M_{\text{min}}$ values to obtain galaxy number densities of $20, 7$ and $2\times10^{-4}\hMpcCube$, respectively.
    }
	\label{tab:hodparams}
	\setlength\tabcolsep{3pt}
	\renewcommand{\arraystretch}{1.0}
	\begin{tabular}{C{4cm}?C{4cm}} \Xhline{3.5\arrayrulewidth}
	Parameter & Value \\ \Xhline{3.5\arrayrulewidth}
	$\log~M_{\text{min}}$ & 11.22, 12.30, 13.22 \\
	$\log~M_0$ & 13.077 \\
	$\log~M_1$ & 14.000 \\
	$\sigma_{\log M}$ & 0.596 \\
	$\alpha$ & 1.0127\\ \Xhline{3.5\arrayrulewidth}
	\end{tabular}
\end{table}

In order to directly compare the performance of the reconstruction method for both halos and galaxies it is necessary to tune the tracer number density to be the same in each case. Unlike the friends-of-friends method which tells us the mass of each halo, the HOD method does not predict galaxy masses and we cannot obtain a given number density by having a galaxy stellar mass cut. We instead tune the galaxy number density by changing the $M_{\text{min}}$ parameter, where $M_{\text{min}}$ is the minimum mass of halos which can host a central galaxy.

The galaxy bias can be calculated in the same way as the halo bias, and also remains constant on large scales.

\section{Results and Discussion}
\label{sec:results}

Fig.~\ref{fig:density_fields} shows a visual comparison of the initial and nonlinear matter density fields, the nonlinear halo field, and the reconstructed density fields from the nonlinear dark matter and halo distributions. All density fields are smoothed using a Gaussian filter with $R=2$\hMpc, with the exception of 1a), which we have left unsmoothed for comparison with 1b). For panels 1a) and 1b), the initial matter density field at $z=49$ has been calculated using TSC mass assignment, and extrapolated to $z=0$ using the $\Lambda$CDM linear growth factor. In the second and third columns we show the nonlinear matter and halo density fields respectively on the top, with the resulting reconstructed density field on the bottom. In panels 1a), 1b), 2b) and 3b) there are some regions where the density contrast $\delta$ is less than $-1$, implying a negative density $\rho$: for 1a) and 1b) this is simply a result of the fact that these fields are linearly extrapolated versions of the initial density field, which is also true to leading order for the reconstructed density fields in 2b) and 3b). Visually, there is a greater similarity between 1b) and 2b) than 1b) and 3b), which is because the halo field contains less information than the dark matter field, in particular on small scales.

To test the performance of our reconstruction method quantitatively, we define the correlation coefficient between two density fields $\delta_1$ and $\delta_2$ as
\begin{equation}
r_{12} = \dfrac{\tilde{\delta}_1\tilde{\delta}_2^* + \tilde{\delta}_1^*\tilde{\delta}_2}{2\sqrt{\tilde{\delta}_1\tilde{\delta}_1^*}\sqrt{\tilde{\delta}_2\tilde{\delta}_2^*}},
\end{equation}
where a * indicates the complex conjugate, and $\tilde{\delta}$ is the Fourier transform of the density field, $\delta({\bf x})$. The correlation coefficient $r_{12}$ describes the similarity between the two density fields. By definition $r_{12} = 1$ if the two fields are identical and $r_{12} = 0$ if they are completely unrelated. 
We are interested in the correlation between the initial and final density fields, which we denote with $r_\text{if}$, and the correlation between the initial and reconstructed density fields, which we denote with $r_\text{ir}$.
We expect to find that $r_\text{if}$ 
is closer to $1$ on large scales where evolution is linear, with a decline towards $0$ on smaller scales where matter has clustered strongly. The performance of the reconstruction method can be quantified by the difference in $r_{\text{if}}$ and $r_{\text{ir}}$, which tells us how much information it has recovered from the initial conditions. As the main aim of this study is to analyse the ability of the reconstruction method to recover the BAO peaks, it is important to observe an improvement on the scales where the first few and most prominent peaks in the power spectrum $P(k)$ occur ($0.05 \lesssim k \lesssim 0.3~h$Mpc$^{-1}$). To assess quantitatively the reconstruction performance in different scenarios, we define $k_{80}$, $k_{50}$ and $k_{20}$ to be the wavenumbers at which the corresponding reconstructed density field is $80\%$, $50\%$ and $20\%$ correlated with the initial conditions, respectively.

\subsection{Comparison of mass assignment schemes}

Before testing the effects of tracer biases, we first compare the different mass assignment methods described in Section~\ref{sec:halos} in order to better understand their impact on reconstruction performance.

\begin{figure}
    \centering
	\includegraphics[width=\linewidth,angle=0]{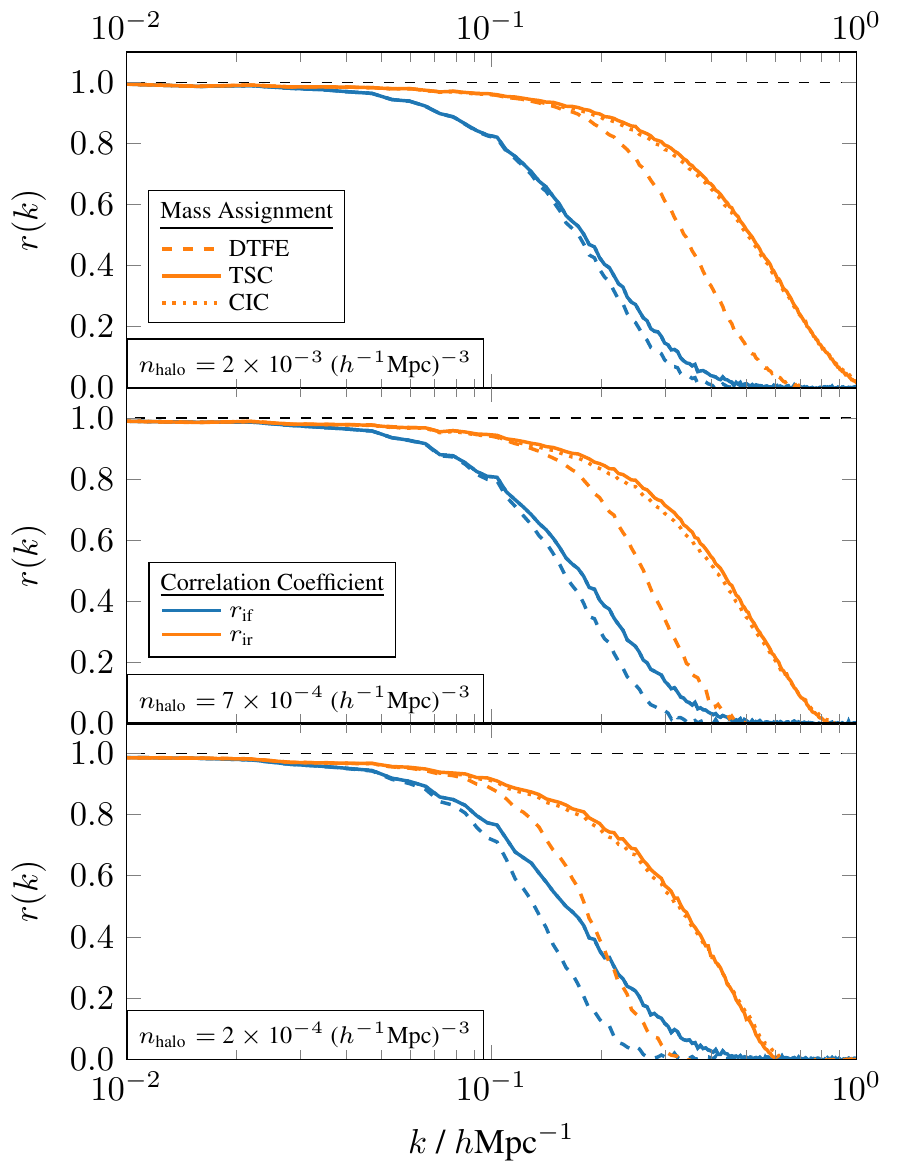}
    \vspace{-4ex}  
	\caption{Correlation coefficients between the initial and final density fields, $r_{\text{if}}$ (blue), and between the initial and reconstructed density fields, $r_{\text{ir}}$ (orange). The final and reconstructed density fields were calculated using the halo number density, $\delta_h$, obtained using the DTFE (dashed curves), TSC (solid) and CIC (dotted) mass assignment schemes. The three panels correspond to different halo number density samples, $n_{\rm halo}=2\times10^{-3}$, $7\times10^{-4}$ and $2\times10^{-4}(h^{-1}{\rm Mpc})^{-3}$ (from top to bottom).
    }
	\label{fig:halotest}
\end{figure}

The result is shown in Fig.~\ref{fig:halotest}, with $k_{80}$, $k_{50}$ and $k_{20}$ values presented in Table~\ref{tab:DTFEvsTSC}. From Fig.~\ref{fig:halotest} it is clear that both the CIC and TSC mass assignments perform better than DTFE mass assignment, with improvements found in both $r_{\text{if}}$ and $r_{\text{ir}}$. Regardless of the method used for mass assignment, we find reconstruction to be more effective when using a high tracer number density, as expected. On the other hand, when moving from DTFE to CIC/TSC mass assignment greater improvements are found when the tracer number density is lower, and in the bottom panel we can see that the nonlinear TSC density field is actually more strongly correlated with the initial conditions than the reconstructed density field from DTFE for $k\gtrsim0.2~h$Mpc$^{-1}$.

The fact that TSC/CIC mass assignment results in a greater improvement over DTFE mass assignment when applied to sparse tracer catalogues is due to the adaptive nature of the DTFE formalism. In DTFE, halos in low density regions distribute their mass up to distances many times the mean halo separation, which effectively corresponds to a large scale smoothing of the density field and inevitably erases information. The largest effective smoothing is for the sparsest halo sample, which is also the one which shows the largest difference in $r_{\text{ir}}$ between the DTFE and the TSC/CIC mass assignments (see Fig.~\ref{fig:halotest}). On the other hand, the performances of TSC and CIC are very similar, with the former slightly better than the latter. Given these tests, in the rest of our analyses we use TSC mass assignment.

\begin{table}
	\centering
	\caption{The wavenumbers $k_{80}$, $k_{50}$ and $k_{20}$ corresponding to the correlation coefficient, $r_{\text{ir}}$, between the initial and reconstructed density fields for two mass assignment schemes, TSC and DTFE, and for three halo samples with different number densities. The wavenumber $k_f$ corresponds to the $k$ value where $r_{\text{ir}}=f$ per cent.
	}
	\setlength\tabcolsep{3pt}
	\renewcommand{\arraystretch}{1.0}
	\begin{tabular}{C{1.5cm}C{1.5cm}?C{1.5cm}C{1.5cm}C{1.5cm}} \Xhline{3.5\arrayrulewidth}
    $n_{\text{halo}}$ & Method & $k_{80}$ & $k_{50}$ & $k_{20}$\\ \Xhline{3.5\arrayrulewidth}
	2 $\times$ 10$^{-3}$ & DTFE & 0.22 & 0.34 & 0.46\\
	& TSC & 0.30 & 0.51 & 0.73\\ \hline
	7 $\times$ 10$^{-4}$ & DTFE & 0.18 & 0.26 & 0.34\\
	& TSC & 0.24 & 0.43 & 0.61 \\ \hline
	2 $\times$ 10$^{-4}$ & DTFE & 0.12 & 0.18 & 0.24\\
	& TSC & 0.18 & 0.33 & 0.47 \\ \Xhline{3.5\arrayrulewidth}
	\end{tabular}
	\label{tab:DTFEvsTSC}
\end{table}

\subsection{Comparison of reconstruction grid resolutions}
\label{sec:rescomp}

Here we investigate the optimal resolution of the regular grid used for the reconstruction procedure. 
Increasing the grid size, that is reducing the grid spacing, allows us to better recover the initial density on small scales and to reduce discretization errors when solving Eq.~(\ref{eq:rec_eqn_bias}) numerically. However, this comes at the price of higher computational resources. There is a physical scale below which structure formation is highly nonlinear, representing a physical limit down to which our method can recover the initial density field. This limiting scale can be reached by using a high number density of tracers, such as when applying the reconstruction to the dark matter distribution, however, in the case of halo or galaxy distributions, the limiting scale is likely higher and arises due to the sparsity of the tracer distribution.
\begin{figure}
    \centering
	\includegraphics[width=\linewidth,angle=0]{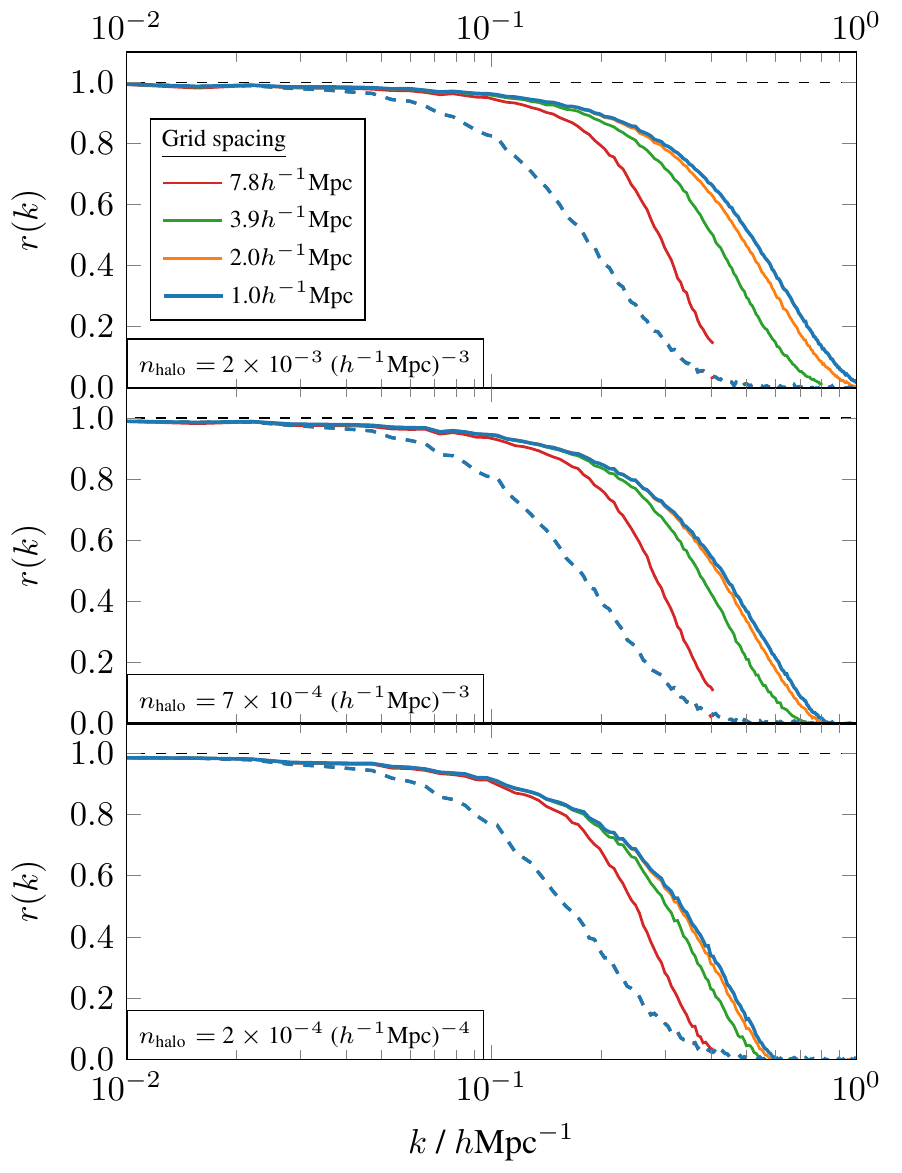}
    \vspace{-4ex}
	\caption{ \changed{Correlation coefficients between the initial and final density fields $r_{\text{if}}$ (dashed), and between the initial and reconstructed density fields $r_{\text{ir}}$ (solid)} for the $n_{\rm halo}=2\times10^{-3}$ (top panel), $7\times10^{-4}$ (middle panel) and $2\times10^{-4}$ (bottom panel) $(h^{-1}$Mpc$)^{-3}$ halo reconstruction performed using different grid resolutions. The legend shows the reconstruction grid cell spacings. Note that the pink solid lines (the results from grid size $128^3$) do not reach $r(k)=0$ because for this grid size the power spectrum is only measured down to a length scale corresponding to $k=0.4h$Mpc$^{-1}$.
    }
	\label{fig:rescomp}
\end{figure}

\changed{We note that varying the grid size employed by our calculation is not the same as varying the smoothing scale used for linear reconstruction methods. Our method is fully non-linear and does not employ smoothing apart from the effective smoothing caused by assigning particles to the computational grid using, e.g., TSC. Starting from a uniform distribution, our reconstruction finds the minimum displacement field needed to obtain the present day mass distribution. Using a smaller grid spacing does not affect the large-scale modes of the displacement field (although it can reduce discretisation errors) and only leads to recovering smaller-scale modes. If the scales are small enough to be affected by nonlinear structure formation, then the recovered small-scale displacement field is uncorrelated with the original field. Thus, decreasing the grid size does not affect our reconstruction. In contrast, the performance of linear reconstruction methods, such as the inverse Zel'dovich approach of \citet{2007ApJ...664..675E}, is sensitive to the employed smoothing scale. This is because that reconstruction procedure is based on an analytic description of structure formation which is valid only down to quasi-linear scales, with the optimal BAO reconstruction corresponding to a smoothing scale ${\sim}10\hMpc$ \citep[see e.g.][]{Vargas-Magana2017}. }

We employ a grid with uniform spatial resolution, using $\left(2^{n_l}\right)^3$ cubic cells, where $n_l$ is an integer. We test 4 cases, with $n_l$ = 7,8,9 and 10 respectively. This paper uses a cubic simulation box with $1~h^{-1}$Gpc side length, therefore these $n_l$ values correspond to a resolution (cubic cell size) of $\ell=7.81, 3.91, 1.95$ and $0.98~h^{-1}$Mpc respectively. Clearly, for larger boxes, larger $n_l$ are needed to achieve the same spatial resolution. For simplicity we consider only halo reconstruction here.

The results are given in Fig.~\ref{fig:rescomp}. Note that the curves representing the $128^3$ grid reconstruction stop at $k\sim0.4\hMpcInv$ because scales smaller than this cannot be sampled on this coarse grid; the same is true for the $256^3$ grid, which does not sample scales smaller than $k\sim0.8\hMpcInv$. We note that the convergence between different grid resolutions depends sensitively on the tracer number density; for example, grid sizes $\geq256^3$ give similar $k_{80}$ for the case of $n_{\rm halo}=2\times10^{-4}~(h^{-1}{\rm Mpc})^{-3}$, but a $256^3$ grid is clearly insufficient for the other two halo number densities. For all three number densities, the $512^3$ and $1024^3$ grids give comparable results, in particular for $k_{80}$ (while for $k_{50}, k_{20}$ the $512^3$ grid has not completely converged yet). It is also notable that $r_{\text{if}}$ is independent of the grid size, which was found not to be the case for DTFE mass assignment (not shown here).

Computing time is not an issue for our reconstruction method. For the $1024^3$ grid resolution, the reconstruction code takes $\sim20$ minutes with $504$ CPUs, using $180$ GB RAM, and it is much faster for lower grid resolutions. On the other hand, as we shall see below, the grid resolution can be important when including nonlinear and nonlocal halo bias in the reconstruction, because a higher resolution means that $\delta_h$ in Eq.~(\ref{eq:rec_eqn_bias}) can become large in cells from high-density regions, and this will effect the reconstruction performance, resulting in a severe constraint on $b_2$, namely $|b_2|\ll1$. To illustrate the impacts of biased halo reconstruction, therefore, in what follows we opt to use the $512^3$ grid for all our tests. In general, however, where computational resources allow, a higher-resolution grid is recommended to make the best of the reconstruction method.

\subsection{Effects of varying tracer bias}
\label{sec:bias}

Having fixed the mass assignment scheme and grid resolution, we now move on to analyse the impacts on the reconstruction performance of varying the tracer bias parameters. We start by varying the linear bias, $b_1$, then proceed to vary the nonlocal bias, $\gamma_2$, and, finally, the nonlinear bias at quadratic order, $b_2$, as described in Section~\ref{sec:method}. More explicitly, we first test a range of values for the linear bias $b_1$, fixing $\gamma_2=b_2=0$, then we fix $b_1$ to the best-fit value and test multiple values of $\gamma_2$, then again we fix both $b_1$ and $\gamma_2$ to their best-fit values and study the effect of varying $b_2$. In this subsection we focus on the correlation coefficients of the reconstructed density fields, with the impact on the BAO peak recovery being studied in the following subsection.

\begin{figure*}
    \centering
	\includegraphics[width=1.03\linewidth,angle=0]{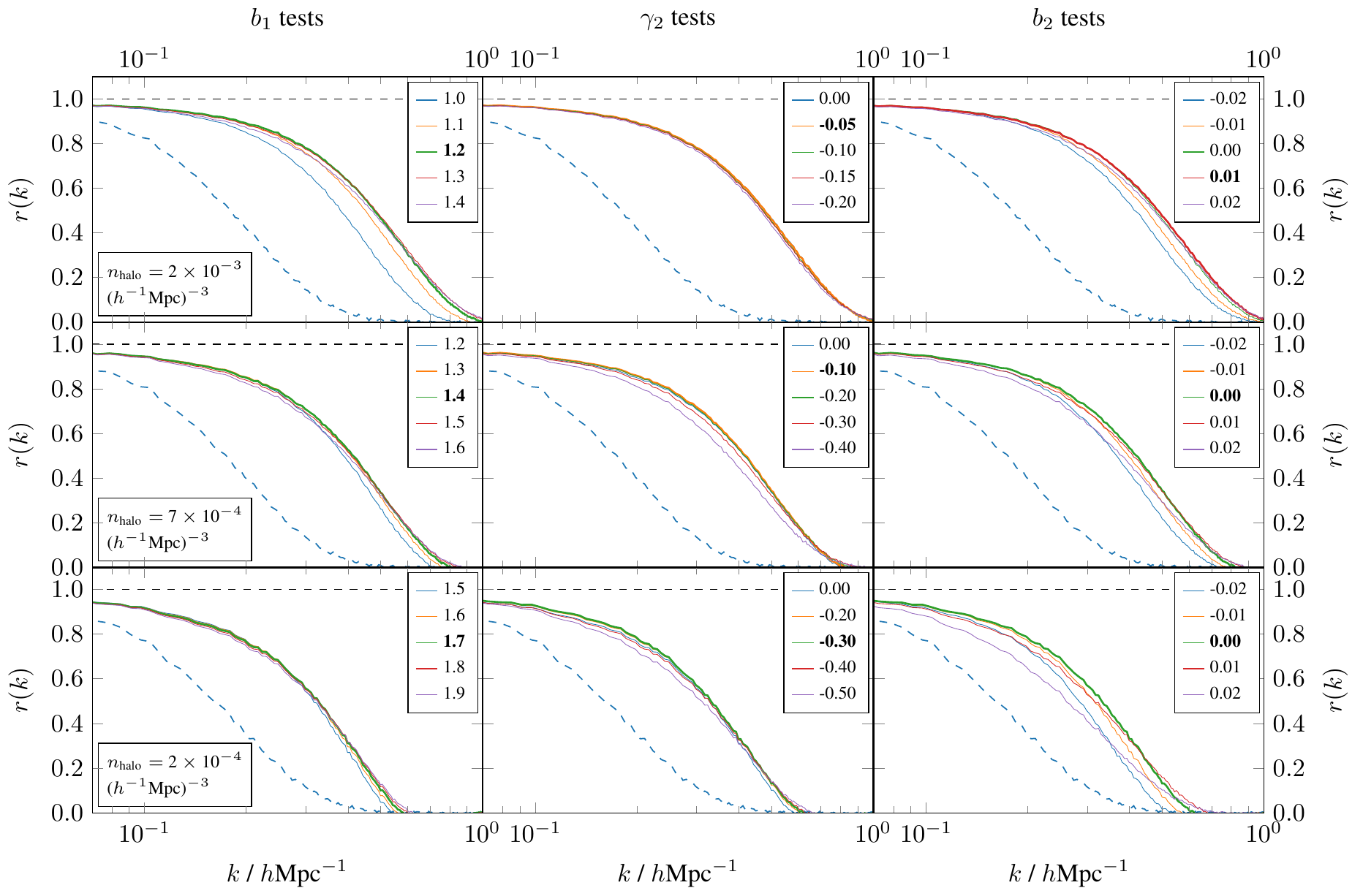}
	\vspace{-4ex}
	\caption{ \changed{Correlation coefficients between the initial and final density fields $r_{\text{if}}$ (dashed), and between the initial and reconstructed density fields $r_{\text{ir}}$ (solid)} for three halo samples with different number densities (each row corresponds to a different number density). Each column tests a different halo bias parameter: the linear bias, $b_1$ (left column); the nonlocal bias, $\gamma_2$, using the optimal $b_1$ value (middle column); and the quadratic bias, $b_2$, using the optimal $b_1$ and $\gamma_2$ values (right column).
	The optimal value in each panel is indicated by a thicker curve and a bold value in the legend.
    }
    \label{fig:halobias}
\end{figure*}

\begin{figure*}
    \centering
	\includegraphics[width=1.03\linewidth,angle=0]{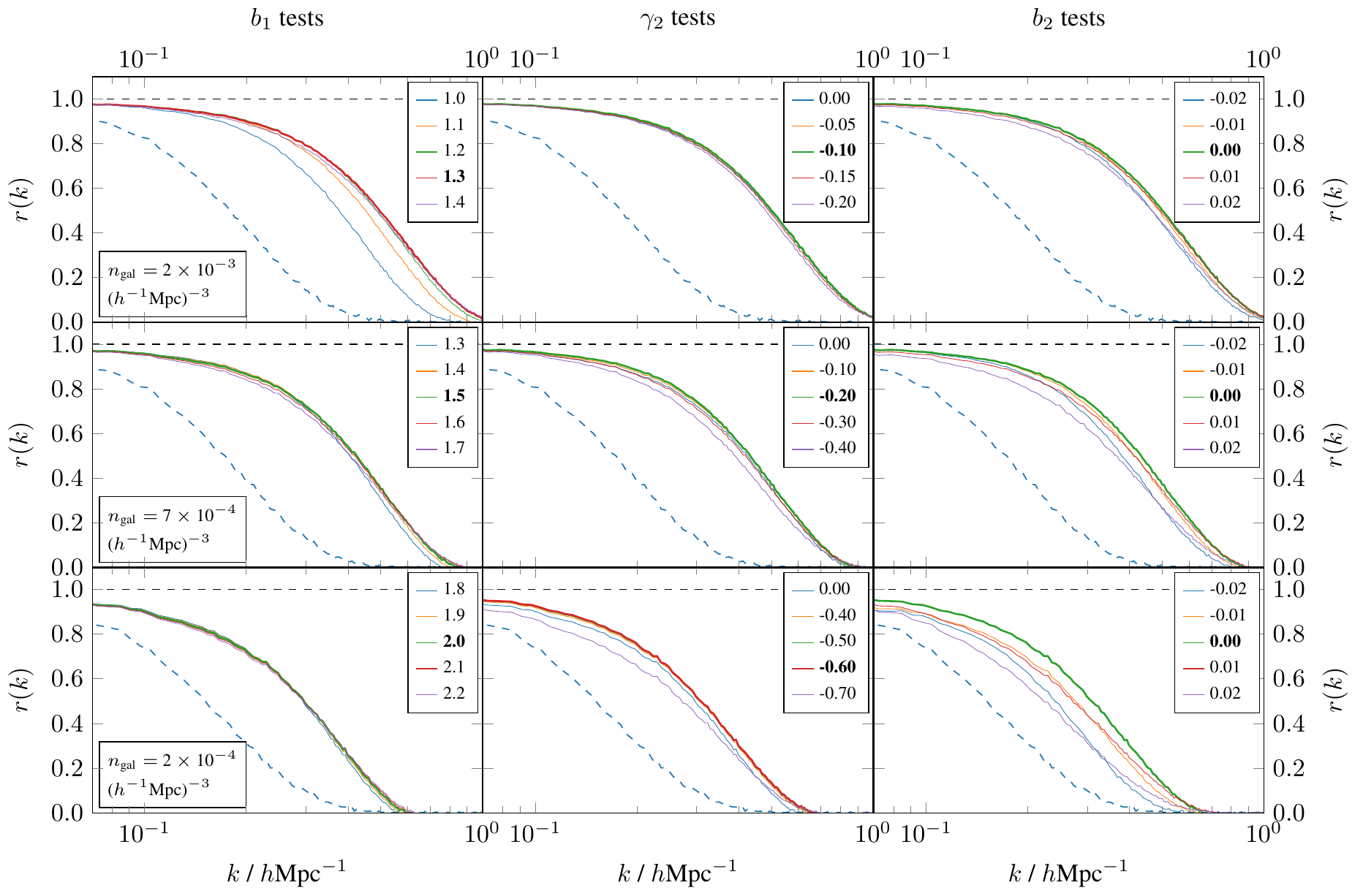}
    \vspace{-4ex}
	\caption{Same as Fig. \ref{fig:halobias}, but now the reconstruction has been performed using HOD galaxy distributions which are built on the halo catalogues used for Fig.~\ref{fig:halobias}. Each row corresponds to a different galaxy number density which is the same as the halo number density in the corresponding row in Fig.~\ref{fig:halobias}.
	}
    \label{fig:galaxybias}
\end{figure*}

Figures \ref{fig:halobias} and \ref{fig:galaxybias} show our findings when applying reconstruction to the halo and galaxy distributions respectively.
All panels show the correlation coefficients between the linear matter and nonlinear tracer density fields ($r_{\text{if}}$; dashed), along with those between the linear matter and reconstructed density fields ($r_{\text{ir}}$; solid) for a range of bias parameter values for $b_1$ (left column), $\gamma_2$ (middle column) and $b_2$ (right column). As the difference between curves is subtle in many cases, we indicate the chosen 'best' configuration by a thicker curve and a bold value in the legend. The $k_{80}$, $k_{50}$ and $k_{20}$ values for the highest and lowest tracer number densities are given in Table~\ref{tab:biask80}, and we refer to this in our analysis. Given the quite similar behaviour seen in Figs.~\ref{fig:halobias} and \ref{fig:galaxybias}, in the discussion below we focus on the case of halo reconstruction, and comment on galaxy reconstruction when appropriate.

As noted above, a common feature in both $r_{\text{if}}$ and $r_{\text{ir}}$ is the decrease of the correlation coefficient from approximately $1.0$ on large scales to $0.0$ on small scales, and the rate of this decrease is slower for higher tracer number densities, which contain more accurate information about the underlying dark matter field. In general, reconstruction boosts the correlation coefficient and extends the range of scales over which it is nonzero. We have tested five values of $b_1$ for each number density, with the central value being the one calculated using the method outlined in Section~\ref{sec:halos}. The measured values are 
$b_1=1.2, 1.4$ and $1.7$ for $n_{\text{halo}}=2\times10^{-3}, 7\times10^{-4}$ and $2\times10^{-4}\hMpcCube$ respectively.
The reconstruction performance is quite sensitive to the value of $b_1$ in the highest number density case, though the range of $b_1\in[1.2,1.3]$ seems to give very similar results. We chose $b_1=1.2$ as our best-fit value to be fixed when varying $\gamma_2$ and $b_2$, despite the fact that $b_1=1.3$ gives slightly better results on small scales ($k>0.6~h$Mpc$^{-1}$), as we are more interested in the large scales when aiming to recover the BAO peaks. We choose $b_1=1.4$ and $1.7$ for $n_{\text{halo}}= 7\times10^{-4}$\hMpcCube and $2\times10^{-4}$\hMpcCube respectively, noting that the optimal $b_1$ value for reconstruction takes the value measured in the simulation for all three number densities. For the two lowest number density samples, the linear bias is not very important and adding $\pm0.2$ does not significantly change the reconstruction performance; in the high number density case, however, the result is more sensitive to $b_1$ but increasing $b_1$ by up to $0.2$ from its best-fit value again has a negligible impact on the correlation coefficient of the reconstructed density field. This is positive news for reconstruction in real observations, where $b_1$ is usually not known accurately.

We next employ the optimal linear bias value $b_1$ for each number density and repeat the reconstruction process by varying the nonlocal bias parameter $\gamma_2$ in the central column of Figs.~\ref{fig:halobias} and \ref{fig:galaxybias}. Applying Eq.~(\ref{eq:gamma2}), we predict $\gamma_2\approx$ $-0.06$, $-0.11$ and $-0.20$ for the three halo catalogues with decreasing number densities; while trying a range of values for $\gamma_2$ in the reconstruction we find $\gamma_2\approx -0.05$, $-0.10$ and $-0.30$, respectively, to be the best values. Although not shown here, using the DTFE mass assignment scheme results in an optimal reconstruction when $\gamma_2\approx-0.2$, $-0.3$ and $-0.5$ for the three corresponding halo number densities.
It is noteworthy that the two mass assignment methods lead to different optimal values of the nonlocal bias, suggesting that the difference in the methods introduces an additional non-physical bias. When using TSC mass assignment the optimal $\gamma_2$ agree more closely with the perturbation theory prediction \citep{2012PhRvD..85h3509C}, although this agreement is worse in the lowest number density case of halo reconstruction and in galaxy reconstruction. Among the three halo number densities, we find that the greatest improvement in reconstruction performance when including nonlocal bias is attained for the sparsest sample, where $n_{\rm halo}=2\times10^{-4}$\hMpcCube, for which $\gamma_2$ is also the largest. Even in this case, the increase of $k_{80}$ is marginal ($0.01$), suggesting that including nonlocal bias in the reconstruction will not substantially improve the recovery of BAO peaks.

The right columns of Figs.~\ref{fig:halobias} and \ref{fig:galaxybias} show the reconstruction results by fixing $b_1$ and $\gamma_2$ to their respective best-fit values for each tracer number density, while varying $b_2$ around $b_2=0$. For all but the case of halo reconstruction where $n_{\rm halo}=2\times10^{-3}$\hMpcCube, we find that the best-fit value is $b_2=0$, and that any significant deviation from this value would quickly downgrade the performance. As mentioned above, this is because $b_2$ enters the reconstruction equation (see Eq.~(\ref{eq:rec_eqn_bias})) through $b_2\delta_h^2$, so that in high density regions where $\delta_h\gg1$ this would lead to a large unphysical contribution to the source of that equation. Physically, the $b_2$ bias term has been introduced as a correction which is valid in the regime $\delta\ll1$, and so should really be used only in the mildly nonlinear regime rather than cases where $\delta_h\gg1$. Indeed, we have explicitly checked that for lower reconstruction grid resolutions, e.g., $128^3$ and $256^3$, $b_2$ takes larger nonzero best-fit values; in those cases adding the nonlinear bias indeed leads to noticeable improvements in the correlation coefficients $r_{\rm ir}$, but at the price that $r_{\rm ir}$ for $b_2=0$ is generally much poorer than the $512^3$ grid case to start with (c.f.~Fig.~\ref{fig:rescomp}). Therefore, at least for the method to model nonlinear bias above, using a high-resolution reconstruction grid removes the necessity or appropriateness to include $b_2$. More complicated treatments, e.g., which first smooth the tracer number density field before doing the reconstruction, might reduce the largest values of $\delta_h$ and therefore allow $b_2$ to be included, but this is beyond the scope of this work.

\begin{table*}
	\centering
	\caption{The $k_{80}$, $k_{50}$ and $k_{20}$ values corresponding to $r_{\text{ir}}$ found for halo and galaxy reconstruction with different halo number densities and different bias parameters.
	}
	\setlength\tabcolsep{3pt}
	\renewcommand{\arraystretch}{1.0}
	\begin{tabular}{C{1.5cm}?C{1.0cm}|C{1.0cm}C{1.0cm}C{1.0cm}?C{1.0cm}|C{1.0cm}C{1.0cm}C{1.0cm}?C{1.0cm}|C{1.0cm}C{1.0cm}C{1.0cm}} \Xhline{3.5\arrayrulewidth}
    $n_{\text{halo}}$ & $b_1$ & $k_{80}$ & $k_{50}$ & $k_{20}$ & $\gamma_2$ & $k_{80}$ & $k_{50}$ & $k_{20}$ & $b_2$ & $k_{80}$ & $k_{50}$ & $k_{20}$\\ \Xhline{3.5\arrayrulewidth}
	& 1.0 & 0.23 & 0.38 & 0.54 & 0.00 & 0.28 & 0.48 & 0.68 & -0.02 & 0.25 & 0.41 & 0.60\\
	\multirow{2}{*}{\shortstack{2 $\times$ 10$^{-3}$ \\ \hMpcCube}} & 1.1 & 0.27 & 0.45 & 0.63 & \textbf{-0.05} & \textbf{0.29} & \textbf{0.48} & \textbf{0.68} & -0.01 & 0.27 & 0.45 & 0.64\\
	& \textbf{1.2} & \textbf{0.28} & \textbf{0.48} & \textbf{0.68} & -0.10 & 0.28 & 0.47 & 0.67 & 0.00 & 0.29 & 0.48 & 0.68\\
	& 1.3 & 0.28 & 0.48 & 0.70 & -0.15 & 0.28 & 0.47 & 0.67 & \textbf{0.01} & \textbf{0.28} & \textbf{0.49} & \textbf{0.70}\\
	& 1.4 & 0.26 & 0.47 & 0.69 & -0.20 & 0.27 & 0.46 & 0.66 & 0.02 & 0.27 & 0.47 & 0.69\\ \hline
	& 1.5 & 0.18 & 0.32 & 0.43 & 0.00 & 0.18 & 0.33 & 0.45 & -0.02 & 0.17 & 0.28 & 0.40\\
	\multirow{2}{*}{\shortstack{2 $\times$ 10$^{-4}$ \\ \hMpcCube}} & 1.6 & 0.18 & 0.33 & 0.44 & -0.20 & 0.19 & 0.33 & 0.46 & -0.01 & 0.18 & 0.31 & 0.44\\ 
	& \textbf{1.7} & \textbf{0.18} & \textbf{0.33} & \textbf{0.45} & \textbf{-0.30} & \textbf{0.19} & \textbf{0.33} & \textbf{0.47} & \textbf{0.00} & \textbf{0.19} & \textbf{0.33} & \textbf{0.47}\\
	& 1.8 & 0.18 & 0.33 & 0.46 & -0.40 & 0.18 & 0.33 & 0.47 & 0.01 & 0.17 & 0.31 & 0.47\\
	& 1.9 & 0.17 & 0.32 & 0.46 & -0.50 & 0.16 & 0.31 & 0.46 & 0.02 & 0.14 & 0.26 & 0.42\\ \Xhline{3.5\arrayrulewidth}
	$n_{\text{galaxy}}$ & $b_1$ & $k_{80}$ & $k_{50}$ & $k_{20}$ & $\gamma_2$ & $k_{80}$ & $k_{50}$ & $k_{20}$ & $b_2$ & $k_{80}$ & $k_{50}$ & $k_{20}$\\ \Xhline{3.5\arrayrulewidth}
	& 1.0 & 0.23 & 0.38 & 0.53 & 0.00 & 0.30 & 0.50 & 0.71 & -0.02 & 0.28 & 0.47 & 0.67\\
	\multirow{2}{*}{\shortstack{2 $\times$ 10$^{-3}$ \\ \hMpcCube}} & 1.1 & 0.28 & 0.45 & 0.63 & -0.05 & 0.30 & 0.50 & 0.72 & -0.01 & 0.30 & 0.49 & 0.69\\
	& 1.2 & 0.30 & 0.49 & 0.69 & \textbf{-0.10} & \textbf{0.30} & \textbf{0.50} & \textbf{0.72} & \textbf{0.00} & \textbf{0.30} & \textbf{0.50} & \textbf{0.72}\\
	& \textbf{1.3} & \textbf{0.30} & \textbf{0.50} & \textbf{0.71} & -0.15 & 0.30 & 0.49 & 0.70 & 0.01 & 0.29 & 0.49 & 0.71\\
	& 1.4 & 0.28 & 0.49 & 0.71 & -0.20 & 0.29 & 0.48 & 0.69 & 0.02 & 0.27 & 0.47 & 0.69\\ \hline
	& 1.8 & 0.16 & 0.29 & 0.42 & 0.00 & 0.16 & 0.30 & 0.43 & -0.02 & 0.13 & 0.24 & 0.35 \\
	\multirow{2}{*}{\shortstack{2 $\times$ 10$^{-4}$ \\ \hMpcCube}} & 1.9 & 0.16 & 0.30 & 0.43 & -0.40 & 0.17 & 0.31 & 0.45 & -0.01 & 0.15 & 0.28 & 0.41\\ 
	& \textbf{2.0} & \textbf{0.16} & \textbf{0.30} & \textbf{0.43} & -0.50 & 0.18 & 0.31 & 0.45 & \textbf{0.00} & \textbf{0.18} & \textbf{0.31} & \textbf{0.46}\\
	& 2.1 & 0.16 & 0.30 & 0.43 & \textbf{-0.60} & \textbf{0.18} & \textbf{0.31} & \textbf{0.46} & 0.01 & 0.14 & 0.27 & 0.43\\
	& 2.2 & 0.15 & 0.30 & 0.44 & -0.70 & 0.13 & 0.27 & 0.42 & 0.02 & 0.11 & 0.22 & 0.36\\ \Xhline{3.5\arrayrulewidth}
	\end{tabular}
	\label{tab:biask80}
\end{table*}

In general, the reconstruction performance varies little between the two types of tracers considered, however we find that HOD galaxies have a larger associated linear and nonlocal bias for a given number density. From the simulation we measure $b_1=1.25, 1.5$ and $2.0$ for $n_{\text{halo}}=2\times10^{-3}, 7\times10^{-4}$ and $2\times10^{-4}\hMpcCube$ respectively, and we find these values to be optimal for reconstruction in the three cases ($b_1=1.25$ was not tested but $b_1=1.3$ was the best value chosen). The tests of nonlocal bias found $\gamma_2\approx -0.10$, $-0.20$ and $-0.60$ to be optimal for reconstruction from the three corresponding number density distributions.  We note that while Eq.~(\ref{eq:gamma2}) gives a poor estimate for $\gamma_2$, it need only be multiplied by a factor of 2 to give agreement with the halo reconstruction results.

\begin{figure*}
    \centering
	\includegraphics[width=1.03\linewidth,angle=0]{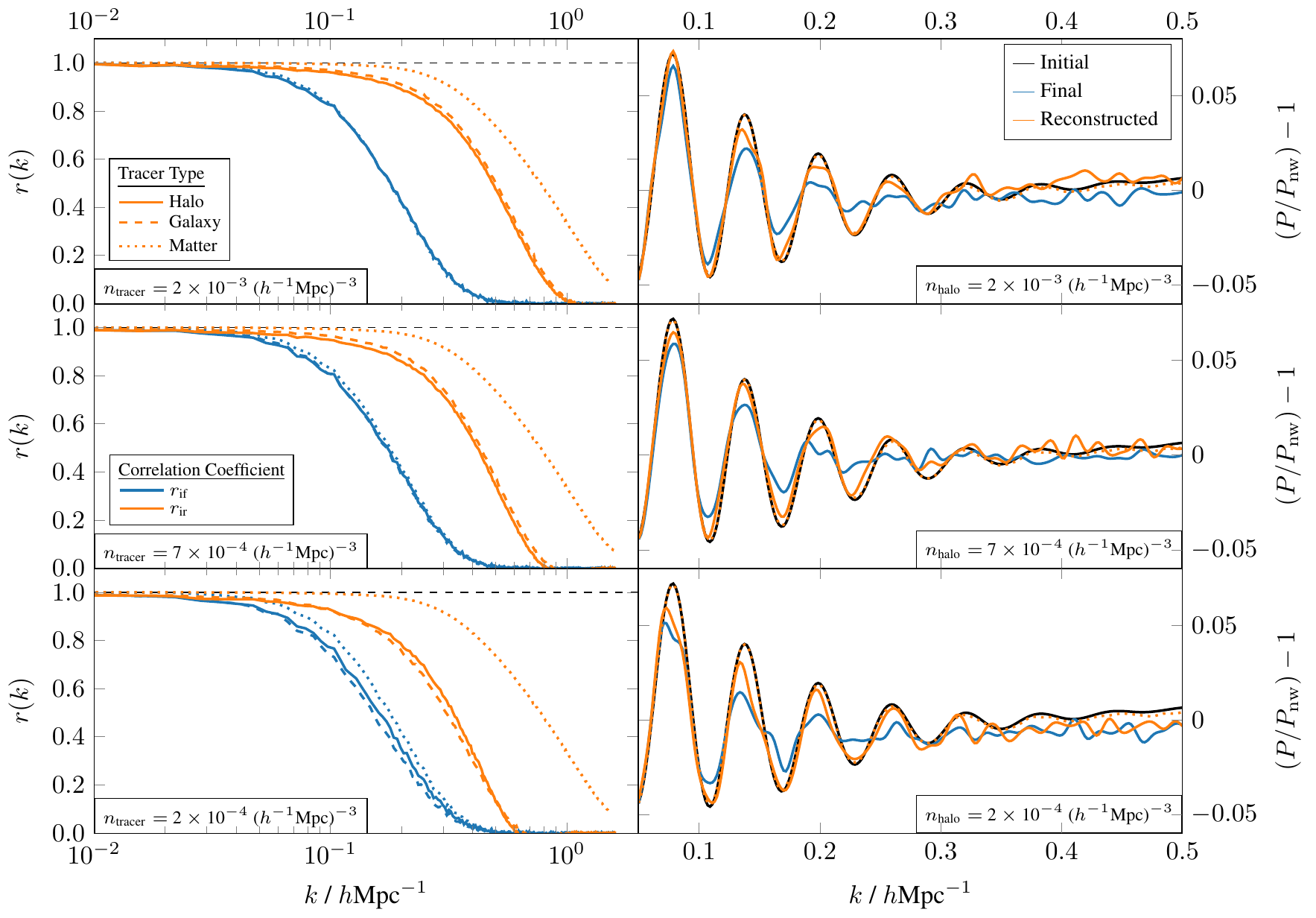}
	\vspace{-4ex}
	\caption{{\it Left column}: correlation coefficients with the initial density distribution of the final ($r_{\text{if}}$; blue) and reconstructed density ($r_{\text{ir}}$; orange) fields, for matter (dotted lines), halos (solid lines) and HOD galaxies (dashed lines). The three panels show the reconstruction results for different tracer number densities, while the dotted lines (for matter reconstruction) are identical in all panels. {\it Right column:} the BAO wiggles in the power spectrum plotted as $P(k)/P_{\rm nw}(k)-1$, where $P(k)$ and $P_{\rm nw}(k)$ are respectively the power spectra extracted from the full and from the no-wiggle simulations. 
	It shows the power spectra ratio for the initial conditions (black solid lines), the final halo distribution (blue solid lines) and the reconstructed initial density from the dark matter (orange dotted lines) and halo (orange solid lines) distributions. For clarity the right-hand panels do not show the result of the HOD galaxy reconstruction. All halo and galaxy reconstruction results are obtained using the optimal bias parameter values $(b_1,\gamma_2,b_2)$ from Figs.~\ref{fig:halobias} and \ref{fig:galaxybias}.
	}
	\label{fig:BAO}
\end{figure*}

\subsection{Recovery of the BAO peaks}
\label{sec:BAO}

Having found the optimal bias values $(b_1,\gamma_2,b_2)$ for each tracer (halo and galaxy) catalogue and number density sample, we now assess the recovery of the BAO peaks using biased tracer reconstruction.

The left panels of Figure \ref{fig:BAO} show the correlation coefficients $r_{\rm if}$ (blue) and $r_{\rm ir}$ (orange) from halo (solid lines), galaxy (dashed lines) and matter (dotted lines) reconstruction for the three tracer number densities as before, decreasing from top to bottom. For matter reconstruction the curves are the same in all three rows. These plots show that tracer reconstruction generally performs worse than matter reconstruction, even for the highest number density used here, but increasing $n_{\rm tracer}$ does bring $r_{\rm ir}$ closer to the matter reconstruction case; it will be interesting to study the value of $n_{\rm tracer}$ at which $r_{\rm ir}$ for tracers and matter become very close. On the other hand, $r_{\rm if}$ depends less sensitively on the tracer number density.

The panels in the right-hand column of Figure \ref{fig:BAO} show the power spectra of the initial, nonlinear halo and reconstructed density fields in the form $(P/P_{\mathrm{nw}})-1$, where $P_{\mathrm{nw}}$ comes from a simulation identical to that of the original, except that there are no BAO wiggles in the linear power spectrum used to generated the simulation initial conditions. Plotting this quantity allows us to clearly visualise the damping, due to nonlinear structure formation, and the recovery, due to reconstruction, of the baryon acoustic oscillations in the power spectrum. The black solid curve, which represents the BAO of the initial linear matter power spectrum, is identical in all three rows, whereas the blue curves, which represent the BAO peaks in the $z=0$ halo power spectrum, are dependent on the halo number density $n_{\rm halo}$. The damping of the BAO wiggles is more severe when $n_{\rm halo}$ is low, and the curves become very noisy, particularly in the lowest number density case. The loss of information from the initial conditions increases with the damping of the BAO wiggles, and this is more significant in the lower halo number density cases where $r_{\text{if}}$ drops off towards zero more rapidly. Similarly, the BAO wiggles are recovered to a great extent when the halo number density is greater, as would be expected from the left panels. Note that we have omitted the galaxy power spectra due to the similarity of the cross correlations with halos for all 3 number densities.

The BAO wiggles from the reconstructed density fields are shown in orange in the right panels of Fig.~\ref{fig:BAO}, with dotted and solid lines representing respectively the results from dark matter and halo reconstruction. The dotted orange lines are the same in all three rows, and they show that dark matter reconstruction is capable of recovering the BAO peaks down to $k\approx0.4~h$Mpc$^{-1}$. Halo reconstruction does not perform as well, as expected, but for all three halo number densities, we still observe a substantial recovery of the BAO wiggles, e.g., compared with the blue curves, in the first four peaks, down to $k\approx0.25~h$Mpc$^{-1}$. The improvement is substantial for all halo number densities. Note that the orange and blue curves have been shifted vertically to align them with the black curve. This is because the same value of $b_1$ was used for both the wiggle and no-wiggle simulations, when in reality the measured values differ by roughly $1\%$, and so taking the ratio of the $P(k)$ and $P_{\mathrm{nw}}(k)$ propagates this error to $\sim2\%$. It is therefore appropriate to shift the curves to provide a clearer comparison.

\changed{To assess the competitiveness of our method, we can compare the enhancement of the BAO feature with results of other reconstruction approaches. The \citet{2017ApJ...847..110Y} study represents a suitable comparison as they have applied their non-linear reconstruction procedure to similar populations of tracers and redshifts as us. For example, comparing our $n_{\text{halo}}=2\times10^{-3}$\hMpcCube{} results with the $n_{\text{halo}}=2.77\times10^{-3}$\hMpcCube{} there, we find that our method performs better in this case. In particular, there the reconstructed density field is approximately $95\%$ and $65\%$ correlated with the initial conditions at $k =$ 0.1\hMpcInv{} and 0.3\hMpcInv{} respectively (see their Fig. 2), whereas we find a correlation coefficient of $95\%$ and $80\%$ for the same $k$ values. We note, however, that this difference is likely due to the different density assignment schemes used -- DTFE there and TSC here (as we have found above, using DTFE causes additional smoothing of the pre-reconstruction density field, which can downgrade the outcome of reconstruction even though it is not related to the reconstruction method itself).}

\section{Summary, discussion and conclusions}
\label{sec:conclusions}

We have tested the nonlinear density reconstruction method introduced by \citet{PhysRevD.97.023505}, applying it to late-time halo and galaxy distributions, to study the potential of recovering BAO peaks from a tracer field, and how this depends on a number of factors including tracer type, tracer number density, mass assignment scheme, reconstruction grid resolution, and tracer bias parameters up to quadratic order. For this, we have developed an extension to the original \citet{PhysRevD.97.023505} method to incorporate nonlocal and nonlinear tracer bias. These terms can be included naturally in the reconstruction equation -- which is a nonlinear partial differential equation that takes the form of the Monge-Ampere equation -- by changing the coefficients and source terms of the equation. The original numerical algorithm still works efficiently when applied to the new equation.

Our results confirm that, as expected, tracer number density plays an important role in determining the performance of reconstruction (which we assess by calculating the correlation coefficient, $r_{\rm ir}$, between the initial and reconstructed density fields), with higher number density tracers giving larger $r_{\rm ir}$ values. The mass assignment scheme used to calculate the tracer density at each position is another important factor for reconstruction performance, with TSC faring significantly better than DTFE and slightly better than CIC for all tracer number densities used. Using a sufficiently high-resolution computational grid for reconstruction is also crucial, and we find that once the grid cell size decreases to $\sim1-2~h^{-1}$Mpc the results converge for all tracer number densities studied here. Reconstruction from HOD galaxy and halo distributions with the same number density give quite similar results.

Of the three bias parameters studied in this work, the linear tracer bias, $b_1$, is by far the most important. For high tracer number densities the reconstruction performance depends sensitively on it, while this dependence is much weaker for low tracer number densities. In all cases, we find that the linear bias parameter measured in the simulation by comparing the clustering of dark matter and halo/galaxy distributions
works best, but using larger (by up to $0.2$) values does not affect the reconstruction significantly. For TSC mass assignment, we find that the nonlocal bias parameter $\gamma_2$ predicted by perturbation theory is close to the values that give rise to the best reconstruction result, but this is not the case when DTFE mass assignment is used, which is another reason why we use TSC in the bias analysis. Including nonlocal bias, however, only marginally improves $r_{\rm ir}$, with the largest improvement found for the lowest  number density sample, for which the optimal $|\gamma_2|$ value is also the largest. Finally, the nonlinear bias at quadratic order, $b_2$, if nonzero, can lead to poorer reconstruction, because our reconstruction method calculates the displacement field on all scales, while the nonlinear bias does not work on small scales where the density field can become large.

These results are confirmed by visually inspecting the recovery of the BAO peaks, as shown in the right panels of Fig.~\ref{fig:BAO}. We can see there that, when applied to halo reconstruction using the best-fit bias parameters, our method substantially improves the recovery of the first few BAO peaks compared with the case of no reconstruction, down to $k\sim0.25~h$Mpc$^{-1}$.

For all the tracer reconstruction results shown here, the tracer density field, $\delta_h$, is calculated by treating the tracers as particles of equal mass, which is a simplified assumption. For example, some halos are more massive (e.g., $>10^{15}h^{-1}M_\odot$) than others (e.g., $<10^{12}h^{-1}M_\odot$). Naturally, more massive halos contain more matter, implying that the nonlinear dark matter field may be more reliably constructed using a mass-weighted halo number density field. To verify this, we have also carried out halo reconstruction tests in which $\delta_h$ is calculated using the actual masses of the haloes. However, this approach leads to a poorer reconstruction, with the resulting $r_{\rm ir}$ being smaller than the ones seen in Figure \ref{tab:DTFEvsTSC}, in particular for the high halo number density case. This happens regardless of the value of $b_1$ used, and it could be because the simple mass-weighting scheme above gives too little weight to low mass halos, which are important tracers of the underlying matter field. This therefore indicates a more sophisticated weight scheme may be required. We leave an investigation on this to future work.

As mentioned above, in principle our method for biased tracer reconstruction can be straightforwardly generalised to higher-order bias terms. For example, the nonlocal bias at cubic order can be similarly expressed in terms of derivatives of the displacement potential, $\theta$, amounting to a further change of various coefficients in the reconstruction equation, Eq.~(\ref{eq:rec_eqn_bias}). However, we have decided not to pursue this line of research, given that the effect of including bias terms up to the quadratic order is already small.

As the first attempt to add more reality to the reconstruction method of \citet{PhysRevD.97.023505}, in this work we have only considered a few simple cases of tracer reconstruction. In order to be able to apply the method to observational data, such as galaxy catalogues, a few important factors need to be taken into account. First, while the tests in this paper have all been done in a cubic box for a fixed snapshot ($z=0$), both the spatial and the redshift distributions of galaxies in real observations are more complicated. For example, observed galaxy catalogues are usually in a lightcone rather than a box, and certain regions of the field of view are masked with no data collected; for reconstruction we will need to embed the lightcone into a cubic box, adding a density field (e.g., zero, random, or uniform) outside the lightcone ensuring periodic boundary conditions. Second, real galaxy catalogues may suffer from incompleteness issues which can be caused by observing conditions, redshift failures, fibre collisions, etc., and care must be taken to deal with this or make corrections. Third, while we have used constant bias values in this study, for observed galaxy catalogues covering a significant redshift interval the bias parameters do evolve, and this should be taken into account as well. Fourth, in this study we have not considered the redshift space distortions of galaxy line-of-sight (los) distances, but in reality only the redshifts of galaxies are measured, whose relation with the los distances are complicated due to coherent and virialised galaxy motions \citep[see, e.g.,][for some recent studies of reconstruction from redshift space]{Zhu:2017vtj,Hada:2018fde}. It will be interesting to extend the reconstruction method used here to include redshift space distortions. It is also useful to apply the method to different tracer types (bright galaxies, luminous red galaxies, emission line galaxies, quasars, etc.), which cover different redshift ranges and have different bias properties. In order to verify its accuracy, it is also important to test the final pipeline using some realistic mock galaxy catalogues \citep[e.g.,][]{Smith:2017tzz}. We leave these interesting developments to future works.

\section*{Acknowledgements}
We thank Xin Wang and Hong-Ming Zhu for helpful discussions during this project \changed{and the anonymous referee for their insightful comments}. JB and BL are supported by the European Research Council (ERC-StG-716532-PUNCA), BL and MC are supported by the STFC through grant ST/P000541/1.
This work used the DiRAC Data Centric system at Durham University, operated by the Institute for Computational Cosmology on behalf of the STFC DiRAC HPC Facility (www.dirac.ac.uk). This equipment was funded by BIS National E-infrastructure capital grant ST/K00042X/1, STFC capital grant ST/H008519/1, and STFC DiRAC Operations grant ST/K003267/1 and Durham University. DiRAC is part of the National E-Infrastructure. 



\bibliographystyle{mnras}
\bibliography{bibliography}

\begin{thebibliography}{}
\makeatletter
\relax
\def\mn@urlcharsother{\let\do\@makeother \do\$\do\&\do\#\do\^\do\_\do\%\do\~}
\def\mn@doi{\begingroup\mn@urlcharsother \@ifnextchar [ {\mn@doi@}
  {\mn@doi@[]}}
\def\mn@doi@[#1]#2{\def\@tempa{#1}\ifx\@tempa\@empty \href
  {http://dx.doi.org/#2} {doi:#2}\else \href {http://dx.doi.org/#2} {#1}\fi
  \endgroup}
\def\mn@eprint#1#2{\mn@eprint@#1:#2::\@nil}
\def\mn@eprint@arXiv#1{\href {http://arxiv.org/abs/#1} {{\tt arXiv:#1}}}
\def\mn@eprint@dblp#1{\href {http://dblp.uni-trier.de/rec/bibtex/#1.xml}
  {dblp:#1}}
\def\mn@eprint@#1:#2:#3:#4\@nil{\def\@tempa {#1}\def\@tempb {#2}\def\@tempc
  {#3}\ifx \@tempc \@empty \let \@tempc \@tempb \let \@tempb \@tempa \fi \ifx
  \@tempb \@empty \def\@tempb {arXiv}\fi \@ifundefined
  {mn@eprint@\@tempb}{\@tempb:\@tempc}{\expandafter \expandafter \csname
  mn@eprint@\@tempb\endcsname \expandafter{\@tempc}}}

\bibitem[\protect\citeauthoryear{{Aghamousa} et~al.,}{{Aghamousa}
  et~al.}{2016}]{2016arXiv161100036D}
{Aghamousa} A.,  et~al., 2016, preprint, \href
  {https://ui.adsabs.harvard.edu/#abs/2016arXiv161100036D} {p.
  arXiv:1611.00036} (\mn@eprint {arXiv} {1611.00036})

\bibitem[\protect\citeauthoryear{{Alam} et~al.,}{{Alam}
  et~al.}{2017}]{2017MNRAS.470.2617A}
{Alam} S.,  et~al., 2017, \mn@doi [\mnras] {10.1093/mnras/stx721}, \href
  {https://ui.adsabs.harvard.edu/#abs/2017MNRAS.470.2617A} {470, 2617}

\bibitem[\protect\citeauthoryear{{Alonso}}{{Alonso}}{2012}]{2012arXiv1210.1833A}
{Alonso} D.,  2012, preprint, \href
  {https://ui.adsabs.harvard.edu/#abs/2012arXiv1210.1833A} {p. arXiv:1210.1833}
  (\mn@eprint {arXiv} {1210.1833})

\bibitem[\protect\citeauthoryear{{Aubourg} et~al.,}{{Aubourg}
  et~al.}{2015}]{2015PhRvD..92l3516A}
{Aubourg} {\'E}.,  et~al., 2015, \mn@doi [\prd] {10.1103/PhysRevD.92.123516},
  \href {https://ui.adsabs.harvard.edu/#abs/2015PhRvD..92l3516A} {92, 123516}

\bibitem[\protect\citeauthoryear{{Behroozi}, {Wechsler}  \& {Wu}}{{Behroozi}
  et~al.}{2013}]{2013ApJ...762..109B}
{Behroozi} P.~S.,  {Wechsler} R.~H.,   {Wu} H.-Y.,  2013, \mn@doi [\apj]
  {10.1088/0004-637X/762/2/109}, \href
  {https://ui.adsabs.harvard.edu/#abs/2013ApJ...762..109B} {762, 109}

\bibitem[\protect\citeauthoryear{Berlind \& Weinberg}{Berlind \&
  Weinberg}{2002}]{Berlind:2001xk}
Berlind A.~A.,  Weinberg D.~H.,  2002, \mn@doi [Astrophys. J.]
  {10.1086/341469}, 575, 587

\bibitem[\protect\citeauthoryear{{Brenier}, {Frisch}, {H{\'e}non}, {Loeper},
  {Matarrese}, {Mohayaee}  \& {Sobolevski{\u{i}}}}{{Brenier}
  et~al.}{2003}]{2003MNRAS.346..501B}
{Brenier} Y.,  {Frisch} U.,  {H{\'e}non} M.,  {Loeper} G.,  {Matarrese} S.,
  {Mohayaee} R.,   {Sobolevski{\u{i}}} A.,  2003, \mn@doi [\mnras]
  {10.1046/j.1365-2966.2003.07106.x}, \href
  {https://ui.adsabs.harvard.edu/#abs/2003MNRAS.346..501B} {346, 501}

\bibitem[\protect\citeauthoryear{{Cautun} \& {van de Weygaert}}{{Cautun} \&
  {van de Weygaert}}{2011}]{2011ascl.soft05003C}
{Cautun} M.~C.,  {van de Weygaert} R.,  2011, {The DTFE public software: The
  Delaunay Tessellation Field Estimator code}, Astrophysics Source Code Library
  (\mn@eprint {ascl} {1105.003})

\bibitem[\protect\citeauthoryear{{Chan}, {Scoccimarro}  \& {Sheth}}{{Chan}
  et~al.}{2012}]{2012PhRvD..85h3509C}
{Chan} K.~C.,  {Scoccimarro} R.,   {Sheth} R.~K.,  2012, \mn@doi [\prd]
  {10.1103/PhysRevD.85.083509}, \href
  {https://ui.adsabs.harvard.edu/#abs/2012PhRvD..85h3509C} {85, 083509}

\bibitem[\protect\citeauthoryear{Cole et~al.}{Cole et~al.}{2005}]{Cole:2005sx}
Cole S.,  et~al., 2005, \mn@doi [Mon. Not. Roy. Astron. Soc.]
  {10.1111/j.1365-2966.2005.09318.x}, 362, 505

\bibitem[\protect\citeauthoryear{{Crocce}, {Pueblas}  \&
  {Scoccimarro}}{{Crocce} et~al.}{2006}]{2006MNRAS.373..369C}
{Crocce} M.,  {Pueblas} S.,   {Scoccimarro} R.,  2006, \mn@doi [\mnras]
  {10.1111/j.1365-2966.2006.11040.x}, \href
  {https://ui.adsabs.harvard.edu/#abs/2006MNRAS.373..369C} {373, 369}

\bibitem[\protect\citeauthoryear{{Croft} \& {Gaztanaga}}{{Croft} \&
  {Gaztanaga}}{1997}]{1997MNRAS.285..793C}
{Croft} R. A.~C.,  {Gaztanaga} E.,  1997, \mn@doi [\mnras]
  {10.1093/mnras/285.4.793}, \href
  {https://ui.adsabs.harvard.edu/#abs/1997MNRAS.285..793C} {285, 793}

\bibitem[\protect\citeauthoryear{Desjacques, Jeong  \& Schmidt}{Desjacques
  et~al.}{2018}]{Desjacques:2016bnm}
Desjacques V.,  Jeong D.,   Schmidt F.,  2018, \mn@doi [Phys. Rept.]
  {10.1016/j.physrep.2017.12.002}, 733, 1

\bibitem[\protect\citeauthoryear{{Eisenstein} \& {Hu}}{{Eisenstein} \&
  {Hu}}{1998}]{1998ApJ...496..605E}
{Eisenstein} D.~J.,  {Hu} W.,  1998, \mn@doi [\apj] {10.1086/305424}, \href
  {https://ui.adsabs.harvard.edu/#abs/1998ApJ...496..605E} {496, 605}

\bibitem[\protect\citeauthoryear{Eisenstein et~al.}{Eisenstein
  et~al.}{2005}]{Eisenstein:2005su}
Eisenstein D.~J.,  et~al., 2005, \mn@doi [Astrophys. J.] {10.1086/466512}, 633,
  560

\bibitem[\protect\citeauthoryear{{Eisenstein}, {Seo}, {Sirko}  \&
  {Spergel}}{{Eisenstein} et~al.}{2007}]{2007ApJ...664..675E}
{Eisenstein} D.~J.,  {Seo} H.-J.,  {Sirko} E.,   {Spergel} D.~N.,  2007,
  \mn@doi [\apj] {10.1086/518712}, \href
  {https://ui.adsabs.harvard.edu/#abs/2007ApJ...664..675E} {664, 675}

\bibitem[\protect\citeauthoryear{{Frisch}, {Matarrese}, {Mohayaee}  \&
  {Sobolevski}}{{Frisch} et~al.}{2002}]{2002Natur.417..260F}
{Frisch} U.,  {Matarrese} S.,  {Mohayaee} R.,   {Sobolevski} A.,  2002, \mn@doi
  [\nat] {10.1038/417260a}, \href
  {https://ui.adsabs.harvard.edu/#abs/2002Natur.417..260F} {417, 260}

\bibitem[\protect\citeauthoryear{Fry \& Gaztanaga}{Fry \&
  Gaztanaga}{1993}]{Fry:1992vr}
Fry J.~N.,  Gaztanaga E.,  1993, \mn@doi [Astrophys. J.] {10.1086/173015}, 413,
  447

\bibitem[\protect\citeauthoryear{Hada \& Eisenstein}{Hada \&
  Eisenstein}{2018}]{Hada:2018fde}
Hada R.,  Eisenstein D.~J.,  2018, ] {10.1093/mnras/sty1203}

\bibitem[\protect\citeauthoryear{Hockney \& Eastwood}{Hockney \&
  Eastwood}{1988}]{Hockney_Eastwood_book}
Hockney R.~W.,  Eastwood J.~W.,  1988, Computer Simulation using Particles.
Institute of Physics Publishing

\bibitem[\protect\citeauthoryear{{Ivezi{\'c}} et~al.,}{{Ivezi{\'c}}
  et~al.}{2008}]{2008arXiv0805.2366I}
{Ivezi{\'c}} {\v{Z}}.,  et~al., 2008, preprint, \href
  {https://ui.adsabs.harvard.edu/#abs/2008arXiv0805.2366I} {p. arXiv:0805.2366}
  (\mn@eprint {arXiv} {0805.2366})

\bibitem[\protect\citeauthoryear{{Jasche} \& {Lavaux}}{{Jasche} \&
  {Lavaux}}{2018}]{Jasche2018}
{Jasche} J.,  {Lavaux} G.,  2018, preprint, \href
  {https://ui.adsabs.harvard.edu/#abs/2018arXiv180611117J} {p.
  arXiv:1806.11117} (\mn@eprint {arXiv} {1806.11117})

\bibitem[\protect\citeauthoryear{{Jasche} \& {Wandelt}}{{Jasche} \&
  {Wandelt}}{2013}]{Jasche2013}
{Jasche} J.,  {Wandelt} B.~D.,  2013, \mn@doi [\mnras] {10.1093/mnras/stt449},
  \href {https://ui.adsabs.harvard.edu/#abs/2013MNRAS.432..894J} {432, 894}

\bibitem[\protect\citeauthoryear{{Kitaura} \& {En{\ss}lin}}{{Kitaura} \&
  {En{\ss}lin}}{2008}]{Kitaura2008}
{Kitaura} F.~S.,  {En{\ss}lin} T.~A.,  2008, \mn@doi [\mnras]
  {10.1111/j.1365-2966.2008.13341.x}, \href
  {https://ui.adsabs.harvard.edu/#abs/2008MNRAS.389..497K} {389, 497}

\bibitem[\protect\citeauthoryear{{Laureijs} et~al.,}{{Laureijs}
  et~al.}{2011}]{2011arXiv1110.3193L}
{Laureijs} R.,  et~al., 2011, preprint, \href
  {https://ui.adsabs.harvard.edu/#abs/2011arXiv1110.3193L} {p. arXiv:1110.3193}
  (\mn@eprint {arXiv} {1110.3193})

\bibitem[\protect\citeauthoryear{{Lavaux}}{{Lavaux}}{2016}]{Lavaux2016}
{Lavaux} G.,  2016, \mn@doi [\mnras] {10.1093/mnras/stv2915}, \href
  {https://ui.adsabs.harvard.edu/#abs/2016MNRAS.457..172L} {457, 172}

\bibitem[\protect\citeauthoryear{{Li}, {Zhao}, {Teyssier}  \& {Koyama}}{{Li}
  et~al.}{2012}]{2012JCAP...01..051L}
{Li} B.,  {Zhao} G.-B.,  {Teyssier} R.,   {Koyama} K.,  2012, \mn@doi [J.
  Cosmo. Astropart. Phys.] {10.1088/1475-7516/2012/01/051}, \href
  {https://ui.adsabs.harvard.edu/#abs/2012JCAP...01..051L} {2012, 051}

\bibitem[\protect\citeauthoryear{{Mohayaee}, {Mathis}, {Colombi}  \&
  {Silk}}{{Mohayaee} et~al.}{2006}]{2006MNRAS.365..939M}
{Mohayaee} R.,  {Mathis} H.,  {Colombi} S.,   {Silk} J.,  2006, \mn@doi
  [\mnras] {10.1111/j.1365-2966.2005.09774.x}, \href
  {https://ui.adsabs.harvard.edu/#abs/2006MNRAS.365..939M} {365, 939}

\bibitem[\protect\citeauthoryear{{Padmanabhan}, {Xu}, {Eisenstein}, {Scalzo},
  {Cuesta}, {Mehta}  \& {Kazin}}{{Padmanabhan}
  et~al.}{2012}]{2012MNRAS.427.2132P}
{Padmanabhan} N.,  {Xu} X.,  {Eisenstein} D.~J.,  {Scalzo} R.,  {Cuesta} A.~J.,
   {Mehta} K.~T.,   {Kazin} E.,  2012, \mn@doi [\mnras]
  {10.1111/j.1365-2966.2012.21888.x}, \href
  {https://ui.adsabs.harvard.edu/#abs/2012MNRAS.427.2132P} {427, 2132}

\bibitem[\protect\citeauthoryear{{Peebles}}{{Peebles}}{1989}]{1989ApJ...344L..53P}
{Peebles} P.~J.~E.,  1989, \mn@doi [\apj] {10.1086/185529}, \href
  {https://ui.adsabs.harvard.edu/#abs/1989ApJ...344L..53P} {344, L53}

\bibitem[\protect\citeauthoryear{Press, Teukolsky, Vetterling  \&
  Flannery}{Press et~al.}{2007}]{Press2007}
Press W.~H.,  Teukolsky S.~A.,  Vetterling W.~T.,   Flannery B.~P.,  2007,
  Numerical Recipes: The Art of Scientific Computing, third edn.
Cambridge University Press

\bibitem[\protect\citeauthoryear{{Schaap} \& {van de Weygaert}}{{Schaap} \&
  {van de Weygaert}}{2000}]{2000A&A...363L..29S}
{Schaap} W.~E.,  {van de Weygaert} R.,  2000, \aap, \href
  {https://ui.adsabs.harvard.edu/#abs/2000A&A...363L..29S} {363, L29}

\bibitem[\protect\citeauthoryear{{Schmittfull}, {Baldauf}  \&
  {Zaldarriaga}}{{Schmittfull} et~al.}{2017}]{2017PhRvD..96b3505S}
{Schmittfull} M.,  {Baldauf} T.,   {Zaldarriaga} M.,  2017, \mn@doi [\prd]
  {10.1103/PhysRevD.96.023505}, \href
  {https://ui.adsabs.harvard.edu/#abs/2017PhRvD..96b3505S} {96, 023505}

\bibitem[\protect\citeauthoryear{{Scoccimarro}}{{Scoccimarro}}{1998}]{1998MNRAS.299.1097S}
{Scoccimarro} R.,  1998, \mn@doi [\mnras] {10.1046/j.1365-8711.1998.01845.x},
  \href {https://ui.adsabs.harvard.edu/#abs/1998MNRAS.299.1097S} {299, 1097}

\bibitem[\protect\citeauthoryear{Shi, Cautun  \& Li}{Shi
  et~al.}{2018}]{PhysRevD.97.023505}
Shi Y.,  Cautun M.,   Li B.,  2018, \mn@doi [Phys. Rev. D]
  {10.1103/PhysRevD.97.023505}, 97, 023505

\bibitem[\protect\citeauthoryear{Smith, Cole, Baugh, Zheng, Angulo, Norberg  \&
  Zehavi}{Smith et~al.}{2017}]{Smith:2017tzz}
Smith A.,  Cole S.,  Baugh C.,  Zheng Z.,  Angulo R.,  Norberg P.,   Zehavi I.,
   2017, \mn@doi [Mon. Not. Roy. Astron. Soc.] {10.1093/mnras/stx1432}, 470,
  4646

\bibitem[\protect\citeauthoryear{Teyssier}{Teyssier}{2002}]{Teyssier:2001cp}
Teyssier R.,  2002, \mn@doi [Astron. Astrophys.] {10.1051/0004-6361:20011817},
  385, 337

\bibitem[\protect\citeauthoryear{{Vargas-Maga{\~n}a}, {Ho}, {Fromenteau}  \&
  {Cuesta}}{{Vargas-Maga{\~n}a} et~al.}{2017}]{Vargas-Magana2017}
{Vargas-Maga{\~n}a} M.,  {Ho} S.,  {Fromenteau} S.,   {Cuesta} A.~J.,  2017,
  \mn@doi [\mnras] {10.1093/mnras/stx048}, \href
  {http://adsabs.harvard.edu/abs/2017MNRAS.467.2331V} {467, 2331}

\bibitem[\protect\citeauthoryear{Wang \& Pen}{Wang \& Pen}{2018}]{Wang:2018ika}
Wang X.,  Pen U.-L.,  2018

\bibitem[\protect\citeauthoryear{{Wang}, {Mo}, {Yang}, {Jing}  \& {Lin}}{{Wang}
  et~al.}{2014}]{Wang2014}
{Wang} H.,  {Mo} H.~J.,  {Yang} X.,  {Jing} Y.~P.,   {Lin} W.~P.,  2014,
  \mn@doi [\apj] {10.1088/0004-637X/794/1/94}, \href
  {http://adsabs.harvard.edu/abs/2014ApJ...794...94W} {794, 94}

\bibitem[\protect\citeauthoryear{{Wang}, {Yu}, {Zhu}, {Yu}, {Pan}  \&
  {Pen}}{{Wang} et~al.}{2017}]{2017ApJ...841L..29W}
{Wang} X.,  {Yu} H.-R.,  {Zhu} H.-M.,  {Yu} Y.,  {Pan} Q.,   {Pen} U.-L.,
  2017, \mn@doi [\apj] {10.3847/2041-8213/aa738c}, \href
  {https://ui.adsabs.harvard.edu/#abs/2017ApJ...841L..29W} {841, L29}

\bibitem[\protect\citeauthoryear{{Weinberg}}{{Weinberg}}{1992}]{1992MNRAS.254..315W}
{Weinberg} D.~H.,  1992, \mn@doi [\mnras] {10.1093/mnras/254.2.315}, \href
  {https://ui.adsabs.harvard.edu/#abs/1992MNRAS.254..315W} {254, 315}

\bibitem[\protect\citeauthoryear{{Yu}, {Zhu}  \& {Pen}}{{Yu}
  et~al.}{2017}]{2017ApJ...847..110Y}
{Yu} Y.,  {Zhu} H.-M.,   {Pen} U.-L.,  2017, \mn@doi [\apj]
  {10.3847/1538-4357/aa89e7}, \href
  {https://ui.adsabs.harvard.edu/#abs/2017ApJ...847..110Y} {847, 110}

\bibitem[\protect\citeauthoryear{Zheng et~al.,}{Zheng
  et~al.}{2005}]{Zheng:2004id}
Zheng Z.,  et~al., 2005, \mn@doi [Astrophys. J.] {10.1086/466510}, 633, 791

\bibitem[\protect\citeauthoryear{{Zheng}, {Coil}  \& {Zehavi}}{{Zheng}
  et~al.}{2007}]{2007ApJ...667..760Z}
{Zheng} Z.,  {Coil} A.~L.,   {Zehavi} I.,  2007, \mn@doi [\apj]
  {10.1086/521074}, \href
  {https://ui.adsabs.harvard.edu/#abs/2007ApJ...667..760Z} {667, 760}

\bibitem[\protect\citeauthoryear{{Zhu}, {Yu}, {Pen}, {Chen}  \& {Yu}}{{Zhu}
  et~al.}{2017}]{2017PhRvD..96l3502Z}
{Zhu} H.-M.,  {Yu} Y.,  {Pen} U.-L.,  {Chen} X.,   {Yu} H.-R.,  2017, \mn@doi
  [\prd] {10.1103/PhysRevD.96.123502}, \href
  {https://ui.adsabs.harvard.edu/#abs/2017PhRvD..96l3502Z} {96, 123502}

\bibitem[\protect\citeauthoryear{Zhu, Yu  \& Pen}{Zhu
  et~al.}{2018}]{Zhu:2017vtj}
Zhu H.-M.,  Yu Y.,   Pen U.-L.,  2018, \mn@doi [Phys. Rev.]
  {10.1103/PhysRevD.97.043502}, D97, 043502

\makeatother
\end{thebibliography}



\appendix


\bsp	
\label{lastpage}

\end{document}